\RequirePackage{ifpdf}
\ifpdf % We are running pdfTeX in pdf mode
\documentclass[pdftex]{sigma}
\else
\documentclass{sigma}
\fi

\newcommand{\Z}{{\mathbb Z}}
\newcommand{\R}{{\mathbb R}}
\newcommand{\J}{{\mathcal J}}
\newcommand{\Jb}{{\overline{\mathcal J}}}
\newcommand{\Rig}{{\rm Rig}}
\newcommand{\Pth}{{\mathcal P}}

\newcommand\bom{\boldsymbol{\omega}}
\newcommand\bol{\boldsymbol{\lambda}}
\newcommand\boa{\boldsymbol{\alpha}}
\newcommand{\m}{{\boldsymbol m}}

\newcommand{\boJ}{{\boldsymbol J}}
\renewcommand{\H}{{\mathcal H}}
\begin{document}

\allowdisplaybreaks

\renewcommand{\thefootnote}{$\star$}

\renewcommand{\PaperNumber}{027}

\FirstPageHeading

\ShortArticleName{Level Set Structure of an Integrable Cellular Automaton}

\ArticleName{Level Set Structure \\ of an Integrable Cellular Automaton\footnote{This paper is a
contribution to the Proceedings of the Workshop ``Geometric Aspects of Discrete and Ultra-Discrete Integrable Systems'' (March 30 -- April 3, 2009, University of Glasgow, UK). The full collection is
available at
\href{http://www.emis.de/journals/SIGMA/GADUDIS2009.html}{http://www.emis.de/journals/SIGMA/GADUDIS2009.html}}}

\Author{Taichiro TAKAGI}

\AuthorNameForHeading{T. Takagi}

\Address{Department of Applied Physics, National Defense Academy,
Kanagawa 239-8686, Japan}
\Email{\href{mailto:takagi@nda.ac.jp}{takagi@nda.ac.jp}}

\ArticleDates{Received October 23, 2009, in f\/inal form March 15, 2010;  Published online March 31, 2010}

\Abstract{Based on a group theoretical setting a sort of
discrete dynamical system is constructed and applied to
a combinatorial dynamical system def\/ined on the set of certain Bethe ansatz related objects
known as the rigged conf\/igurations.
This system is then used to study
a one-dimensional periodic cellular automaton
related to discrete Toda lattice.
It is shown for the f\/irst time that the level set of this cellular automaton
is decomposed into connected components and every such component is a torus.}


\Classification{82B23; 37K15; 68R15; 37B15}

\renewcommand{\thefootnote}{\arabic{footnote}}
\setcounter{footnote}{0}

\section{Introduction}\label{sec:1}
The Liouville's theorem on completely integrable systems
is one of the most fundamental results in classical mechanics.
By V.I.~Arnold's formulation one of the claims in the theorem says that any compact connected
level set of completely integrable systems with $N$ degrees of freedom is
dif\/feomorphic to an $N$-dimensional torus~\cite{A}.
It is called an invariant torus and the theorem also claims that
the phase f\/low with the Hamiltonian function determines a conditionally periodic motion on it.

In this paper we study the level set structure of a one-dimensional cellular automaton known as
the periodic box-ball system (pBBS)~\cite{YT, YYT} and construct its invariant tori.
It is {one of the} (ultra-)discrete dynamical systems associated with
integrable non-linear evolution equations.
See the ``picture''
in Subsection~\ref{subsec:4_1}, just below Remark~\ref{rem:may19_1}, for an example of the time evolution of this system, in which
mutually interacting solitons are traveling along it.
This system is attracting attentions because of its relations with discrete Toda lattice \cite{IT},
Bethe ansatz of integrable quantum spin chains~\cite{KTT}, tropical geometry~\cite{InT} and ultradiscrete
Riemann theta functions~\cite{KS1}.

The purpose of this paper is twofold.
The f\/irst is to construct
a discrete dynamical system through a group theoretical setting
which has potentially several applications in (ultra-)discrete integrable systems (Theorem~\ref{th:apr10_1}).
The second is
to make the structure of the level set of pBBS
perfectly clear as one of its applications (Theorem~\ref{th:main}).
It is shown for the f\/irst time that the level set is decomposed into connected components and every such component is a torus,
as if it were a compact level set of completely integrable systems in Hamiltonian mechanics.

Let us begin by describing backgrounds and motivations on the problem more precisely.
Consider
the completely integrable systems with $N$ degrees of freedom again.
Such a system has~$N$ independent f\/irst integrals.
Its level set is determined by f\/ixing the values of the f\/irst integrals, and becomes
an $N$-dimensional manifold in the $2N$-dimensional phase space.
Inspired by this picture in Hamiltonian mechanics,
A.~Kuniba, A.~Takenouchi and the author
introduced the notion of the level set of pBBS~\cite{KTT}, and
provided its volume formula.
In fact it was exactly equal to the same formula
for the enumeration of the of\/f-diagonal solutions to the string center equations in combinatorial
Bethe ansatz~\cite{KN}.
It is based on the notion of rigged conf\/igurations~\mbox{\cite{KKR,KR}}, and is described in the following way.
Let $m_j$ be the number of solitons with amplitude~$j$.
In what follows $m_j$ are supposed to be positive for $1 \leq j \leq N$ and to be zero for $j > N$ for simplicity.
One can think of $\m=(m_j)_{1 \leq j \leq N}$ as a collection of f\/ixed values of the f\/irst integrals.
We denote by $\J (\m)$ the level set specif\/ied by $\m$
and by $\Omega(\m)$
its volume.
Then we have \cite{KTT}
\begin{equation}
\Omega(\m) =
(\det F)\prod_{1 \leq j \leq N} \frac{1}{m_j}
\binom{p_j + m_j - 1}{m_j - 1}.\label{eq:oct10_1}
\end{equation}
Here $p_j$ are positive integers and $F$ is an $N \times N$ matrix with integer entries,
which are explicitly expressed in terms of $\m$ and the system size $L$.

{It can be shown that}
if there is no solitons of common amplitudes then
the level set $\J (\m)$ becomes a torus $F \Z^N \backslash \Z^N $.\footnote{More precisely, it is not a torus but a set of all integer points on the torus $F \Z^N \backslash \R^N $.
For the sake of simplicity we call such a set a torus.
Besides, in this paper we write all the quotient sets as left quotient ones.}
This fact is suggested by
the volume formula (\ref{eq:oct10_1}) that reduces to $\Omega(\m) =\det F$ when $m_j=1$ for every $j$.
However this simple picture fails when there are multiple solitons of common amplitudes.
{If so,} the right hand side of (\ref{eq:oct10_1}) can not be regarded as the volume of a torus any more.
{An expression for} the level set in this generalized case is given by
\begin{equation}\label{eq:jv}
\J (\m) =  A \Z^{m_1+\cdots + m_{N}} \backslash ({\mathcal I}_{m_1}\times
\cdots \times {\mathcal I}_{m_N}),
\end{equation}
where ${\mathcal I}_n = {\mathfrak S}_n \backslash (\Z^n-\Delta_n)$ is
the $n$ dimensional lattice without the diagonal
points $\Delta_n = \{(z_1,\ldots, z_n) \in \Z^n
\mid z_\alpha = z_\beta \hbox{ for some }
1 \le \alpha \neq \beta \le  n \}$
and identif\/ied under the permutations~${\mathfrak S}_n$.
The $A$ is a $\gamma \times \gamma$ symmetric matrix ($\gamma = m_1+\cdots + m_N$) with integer entries \cite{KTT}.

In this paper we show that the level set (\ref{eq:jv}) has generally many connected components, and every such component
is written as $F^{(\boa)} \Z^N \backslash \Z^N $
where $F^{(\boa)}$ is an $N \times N$ matrix
determined by the above $F$ and the symmetry $\boa$ of the system that depends on the initial conditions.
It is also shown that the time evolutions of the cellular automaton yield straight motions on the torus,
just like the phase f\/lows on the invariant torus generated by f\/irst integrals.

The layout of the rest of this paper is as follows.
In Section~\ref{sec:2} we begin with an abstract group theoretical setting and
present a general result,
and one of its specializations called a~direct product model.
Here we establish our f\/irst main result, Theorem~\ref{th:apr10_1}.
This model is interpreted as a discrete dynamical system.
In Section~\ref{sec:3} we construct a specif\/ic example of the direct product model
associated with the rigged conf\/igurations.
We give a review on the periodic box-ball system in Section~\ref{sec:4}, and then
construct its invariant tori based on the results in previous sections.
Our second main result is Theorem~\ref{th:main}.
Properties of the cellular automaton
that follows from this result are discussed in Section~\ref{sec:5}.
Two elementary lemmas are given in Appendix~\ref{app:A}, and
an algorithm for calculating rigged conf\/igurations is presented in Appendix~\ref{app:B}.

\section{Construction of a discrete dynamical system}\label{sec:2}

\subsection{Group theoretical setting}\label{subsec:2_1}

Let $X$ be a set, ${\frak S}(X)$ be the group of all bijections from $X$ to itself.
Then ${\frak S}(X)$ acts on $X$ from the left by $\sigma \cdot x = \sigma (x)$
$(\forall \, x \in X$, $\forall \, \sigma \in {\frak S}(X))$.
{}From now on we assume that every group action is left action
and omit the word ``from the left''.
Given any group $G$ acting on $X$ there is an equivalence relation associated with its action.
Its equivalence classes are called $G$-orbits on $X$.
If there is only one $G$-orbit, the action of $G$ is called \textit{transitive}.

The $G$-orbit containing $x \in X$ is denoted $G \cdot x$, and is
called the $G$-orbit of $x$.
The set of all $G$-orbits on $X$ is denoted $G \backslash X$.
Though one can think of $G \cdot x$ either as an element of $G \backslash X$ or as a subset of $X$,
we shall adopt the latter interpretation and
elements of $G \backslash X$ will be written as~$[x]_G$.
The map $X \ni x \mapsto [x]_G \in G \backslash X$ is called the \textit{canonical map}
associated with $G \backslash X$.
There is a transitive action of
$G$ on $G \cdot x (\subset X)$.
The following lemma is elementary.

\begin{lemma}\label{lem:jan29_1}
Let $G$ and $H$ be any groups that act on $X$ commutatively.
Then $G$ naturally acts on $H \backslash X$.
Namely their is a unique action of $G$ on $H \backslash X$ that is commutative with
the canonical map $X \ni x \mapsto [x]_H \in H \backslash X$.
\end{lemma}

\begin{proof}
Given any $p \in H \backslash X$ one can write it as $p = [x]_H$ for some $x \in X$.
We def\/ine $\varGamma : G \times H \backslash X \rightarrow H \backslash X$ to be a map given
by the relation $\varGamma (g,  p) = [g \cdot x]_H$ for all $g \in G$.
It is easy to see that this map is well-def\/ined and yields the desired action of $G$ by $g \cdot p = \varGamma (g,  p)$.
\end{proof}
Let $G_1$ and $G_2$ be subgroups of ${\frak S}(X)$ that act on $X$ commutatively.
Then $G_1 \cap G_2$ is also a~subgroup of ${\frak S}(X)$
and its action on $X$ is commutative with those of $G_1$ and $G_2$.
By Lemma~\ref{lem:jan29_1} there is a natural action of $G_1$ on $G_2 \backslash X$.
Let $x$ be any element of $X$.
{Then $G_1 \cdot [x]_{G_2}$ is a subset of $G_2 \backslash X$ where
$G_1$ acts on transitively.
Similarly $G_1 \cdot x$ is a subset of $X$ where $G_1$ and $G_1 \cap G_2$ act on commutatively.}
Hence by Lemma~\ref{lem:jan29_1}  there is a natural action of $G_1$ on
$(G_1 \cap G_2) \backslash (G_1 \cdot x)$.
\begin{proposition}\label{prop:jan29_3}
Let $G_1$ and $G_2$ be subgroups of ${\frak S}(X)$ and suppose their actions on $X$ is commutative.
Suppose we have $(G_1 \cdot x) \cap (G_2 \cdot x) \subset (G_1 \cap G_2) \cdot x$ for some $x \in X$.
Then there is a~bijection between $G_1 \cdot [x]_{G_2}$ and $(G_1 \cap G_2) \backslash (G_1 \cdot x)$
that is commutative with the action of $G_1$.
\end{proposition}
\begin{proof}
Let $z$ be any element of $G_1 \cdot [x]_{G_2}$.
Then there exists $g \in G_1$ such that $z = g \cdot [x]_{G_2}$.
Def\/ine $\varXi: G_1 \cdot [x]_{G_2} \rightarrow (G_1 \cap G_2) \backslash (G_1 \cdot x)$
by $\varXi (z) = [g \cdot x]_{G_1 \cap G_2}$.

\textit{Well-definedness:} Suppose there is another $g' \in G_1$ such that $z = g' \cdot  [x]_{G_2}$.
Since the actions of $G_1$ and $G_2$ are commutative we have
$[g \cdot x]_{G_2} =z = [g' \cdot x]_{G_2}$.
Hence there exists $h \in G_2$ such that $h \cdot (g \cdot x) = g' \cdot x$.
This implies $h \cdot x = (g^{-1} \circ g') \cdot x \in (G_1 \cdot x) \cap (G_2 \cdot x) \subset (G_1 \cap G_2) \cdot x$.
Hence one can take $h$ as an element of $G_1 \cap G_2$.
Thus $[g' \cdot x]_{G_1 \cap G_2} = [g \cdot x]_{G_1 \cap G_2}$.

\textit{Commutativity with $G_1$ action:}
Let $g' \in G_1$.
Then $g' \cdot z = (g' \circ g) \cdot [x]_{G_2}$.
Hence $\varXi(g' \cdot z) = [(g' \circ g) \cdot x]_{G_1 \cap G_2} =
[ g' \cdot (g \cdot x)]_{G_1 \cap G_2} =
g' \cdot ([g \cdot x]_{G_1 \cap G_2}) = g' \cdot \varXi (z)$.

\textit{Injectivity:} Suppose $\varXi (z) = \varXi (z')$ for $z,z' \in G_1 \cdot [x]_{G_2}$.
Then there exist $g, g' \in G_1$ such that $z = g \cdot [x]_{G_2}$ and $z' = g' \cdot [x]_{G_2}$
with the property $[g \cdot x]_{G_1 \cap G_2} = [g' \cdot x]_{G_1 \cap G_2}$.
Hence there exists $h \in G_1 \cap G_2$ such that $h \cdot (g \cdot x) = g' \cdot x$.
Then $g \cdot ([x]_{G_2}) = [g \cdot x]_{G_2} = [h \cdot (g \cdot x)]_{G_2} = [g' \cdot x]_{G_2} = g' \cdot ([x]_{G_2})$.
Hence $z=z'$.

\textit{Surjectivity:} Choose an arbitrary element $C$ of $(G_1 \cap G_2) \backslash (G_1 \cdot x)$.
Then there exists $y \in G_1 \cdot x$ such that $C = [y]_{G_1 \cap G_2}$.
Since $y$ lies in $G_1 \cdot x$ there exists $g \in G_1$ such that $y = g \cdot x$.
Let $z = g \cdot [x]_{G_2} \in G_1 \cdot [x]_{G_2}$.
Then $\varXi (z) = [g \cdot x]_{G_1 \cap G_2} = [y]_{G_1 \cap G_2} =C$.
\end{proof}

Among the assumptions of Proposition~\ref{prop:jan29_3}, the condition $(G_1 \cdot x) \cap (G_2 \cdot x) \subset (G_1 \cap G_2) \cdot x$
is rather specif\/ic.
In the next subsection we construct an example of the triplet $(G_1,G_2,X)$ that satisf\/ies this condition for all $x \in X$.

\subsection{Direct product model}\label{subsec:2_2}

Let $X_1$, $X_2$ be sets and $H_1$, $H_2$ groups that act on $X_1$, $X_2$ respectively.
Choose an arbitrary subgroup $H_2'$ of $H_2$.
We def\/ine
\begin{equation}\label{eq:apr21_1}
X_2' = \{ y \in X_2 \mid g \cdot y = y \mbox{ if and only if } g \in H_2' \}.
\end{equation}
In other words $X_2'$ is the set of all elements of $X_2$ whose stabilizer
associated with the $H_2$ action is $H_2'$.

In what follows $H_2$ is supposed to be abelian, and we denote the multiplication as a sum $g+h \in H_2$
for any $g, h \in H_2$.
Then $X_2'$ is invariant under the action of $H_2$.

\begin{remark}\label{rem:apr21_2}
Suppose there are several dif\/ferent actions of the group $G$ on the set $X$.
To distinguish them we write
$\varGamma : G \times X \rightarrow X$ for example to denote
a specif\/ic action of~$G$ on~$X$.
For any such $\varGamma$
there is a
group homomorphism $\rho_{\varGamma} : G \rightarrow {\frak S}(X)$ such that
the relation $\varGamma (g,x) = \rho_{\varGamma} (g) \cdot x$ holds
for all $g \in G, x \in X$.
Such group homomorphisms are called {\em permutation representations}.
\end{remark}

Let $X = X_1 \times X_2'$.
We def\/ine $\varGamma_1 : H_1 \times X \rightarrow X$
to be a diagonal action of
$H_1$ on $X$
in which its action on $X_2'$ is trivial.
In other words we have
$\varGamma_1(g, x) = (g \cdot x_1, x_2)$ for all $g \in H_1, x =(x_1,x_2) \in X$.

Suppose there exists a map $\varGamma_2 : H_2 \times X \rightarrow X$
that has the following property.
For all $g \in H_2, x =(x_1,x_2) \in X$ we have
$\varGamma_2(g, x) = (\varphi (g, x_1,x_2), g \cdot x_2)$ where
$\varphi : H_2 \times X_1 \times X_2' \rightarrow X_1$ is a map
satisfying the relation $\varphi (g + h, x_1,x_2) = \varphi (h , \varphi (g, x_1,x_2),g \cdot x_2)$ for all $g,h \in H_2$.
Then~$\varGamma_2$ yields an action of $H_2$ on $X$.

By def\/inition we have
$\varGamma_2(g, x) = (\varphi (g, x_1,x_2), x_2)$ for all $g \in H_2'$.
This implies that $\varGamma_2 |_{H_2' \times X}$ yields an action of $H_2'$ on $X$ in which its action on $X_2'$ is trivial.

Let $G_1 = \rho_{\varGamma_1} (H_1), G_2 = \rho_{\varGamma_2} (H_2)$, and $G_2' = \rho_{\varGamma_2} (H_2')$
where $\rho_{\varGamma_1}$ and $\rho_{\varGamma_2}$ are the permutation representations.
Note that $G_1$ and $G_2$
depend on the choice of $H_2'$.

In what follows the action of $H_1$ on $X_1$ is supposed to be transitive.
\begin{lemma}\label{lem:feb9_1}
$G_2' \subset G_1 \cap G_2$.
\end{lemma}

\begin{proof}
By def\/inition we have $G_2' \subset G_2$.
The inclusion $G_2' \subset G_1$ holds since the action of $G_2'$
on~$X_2'$ is trivial, and the action of~$G_1$ on~$X_1$ is transitive.
\end{proof}

\begin{lemma}\label{lem:feb9_2}
The following relation holds
\begin{displaymath}
G_2' \cdot x = (G_1 \cap G_2) \cdot x =
(G_1 \cdot x) \cap (G_2 \cdot x),
\end{displaymath}
for all $x \in X= X_1 \times X_2'$.
\end{lemma}
\begin{proof}
By Lemma \ref{lem:feb9_1} we have
\begin{displaymath}
G_2' \cdot x \subset (G_1 \cap G_2) \cdot x \subset
(G_1 \cdot x) \cap (G_2 \cdot x),
\end{displaymath}
for all $x \in X$,
where the latter inclusion is obvious.
The opposite inclusions are proved as follows.
Take any $x = (x_1, x_2) \in X$.
Choose an arbitrary element $y$ of $(G_1 \cdot x) \cap (G_2 \cdot x)$.
Since $y$ lies in $G_2 \cdot x$ there exists $g \in H_2$ such that
$y = \rho_{\varGamma_2} (g) \cdot x = (\varphi (g, x_1,x_2), g \cdot x_2)$.
Then since $y$ lies in $G_1 \cdot x$ this implies $g \cdot x_2 = x_2$, forcing
$g$ to be an element of $H_2'$ by (\ref{eq:apr21_1}).
Hence $y$ is an element of $G_2' \cdot x$.
\end{proof}

In what follows the actions of $G_1$ and $G_2$ on $X$ are supposed to be commutative.

By Proposition~\ref{prop:jan29_3} and Lemma~\ref{lem:feb9_2}
there is a bijection between $G_1 \cdot [x]_{G_2}$ and $G_1 \cap G_2 \backslash (G_1 \cdot x)$
that is commutative with the action of $G_1= \rho_{\varGamma_1} (H_1)$.
Note that Lemma \ref{lem:feb9_2} implies $G_2' = G_1 \cap G_2$.
Note also that for any $x =(x_1,x_2) \in X$ we have
$G_1 \cdot x = X_1 \times \{ x_2 \} \cong X_1$ as a subset of $X$, for
the action of $G_1$ on $X_1$ is transitive and that on $X_2$ is trivial.
Thus we have the following.

\begin{theorem}\label{th:apr10_1}
Let the sets $X_1$, $X_2'$,
the groups $G_1$, $G_2$, $G_2'$, and their actions on the set $X= X_1 \times X_2'$ be defined as above, and $x$ be any element of $X$.
Then there is a bijection between $G_1 \cdot [x]_{G_2}$ and $G_2' \backslash X_1$ that is commutative with the action of $G_1$.
\end{theorem}

This is our f\/irst main result in this paper.

\subsection{Interpretation as a dynamical system}\label{subsec:2_3}

In this subsection we present, without mathematical rigor, an interpretation of the direct product model
as a dynamical system.
It is intended to help readers to have some intuitive physical images on the model.

Consider any completely integrable system with $N$ degrees of freedom in Hamiltonian mechanics.
There are $N$ independent f\/irst integrals.
By f\/ixing their values we obtain the level set, an $N$-dimensional manifold in the phase space $\mathbb{R}^{2N}$.
On the level set there are $N$ mutually commuting phase f\/lows associated with the f\/irst integrals.
Note that the time evolution of the system is one of the phase f\/lows since the Hamiltonian itself is one of the f\/irst integrals.

One can think of the triplet $(G_1,G_2,X)$ in Subsection \ref{subsec:2_2} as a discrete analogue of such a completely integrable system.
Let $\tilde{X} := X_1 \times X_2$ be a non-compact level set and suppose
the groups $G_1$, $G_2$ are acting on $\tilde{X}$ commutatively.
We regard the group action of $G_2$ as the symmetry of the system and assume that it yields
a compact (but not necessarily connected) level set $G_2 \backslash \tilde{X}$.
Then each state of the system
is represented by a $G_2$-orbit $[x]_{G_2}$ for some $x \in \tilde{X}$.
We regard $G_1$ as the group generated by the mutually commuting phase f\/lows associated with the f\/irst integrals.
Then one can think of
$G_1 \cdot [x]_{G_2} (\subset G_2 \backslash \tilde{X})$ as a connected component of the compact level set
$G_2 \backslash \tilde{X}$ containing the state $[x]_{G_2}$.

Suppose the group $G_2$ acts not only on $\tilde{X}$ but also on $X_2$.
Choose an arbitrary subgroup $G_2'$ of $G_2$ and denote by $X_2'$ the set of all elements of $X_2$ whose stabilizer
associated with the $G_2$ action is $G_2'$.
Let $x$ be any element of $X = X_1 \times X_2'$.
Now the model in Subsection~\ref{subsec:2_2}
can be interpreted as follows.
There is a bijection between the connected component of the level set
$G_1 \cdot [x]_{G_2}$
and the quotient set $G_2' \backslash X_1$ that is commutative with the phase f\/lows of the system.
In particular the time evolution of the system is commutative with this bijection.

Consider the completely integrable system in Hamiltonian mechanics again.
Suppose the level set is compact and connected.
Then by Arnold--Liouville theorem it must be dif\/feomorphic to an $N$-dimensional torus~\cite{A}.
It is called an \textit{invariant torus}.
In the following sections we construct a direct product model
with $X_1 = \Z^N$ and $G_2'$ being a sub-lattice of $\Z^N$
which acts on~$X_1$ by left translation.
Now the quotient set $G_2' \backslash X_1$ is, roughly speaking, a discrete
analogue of invariant torus.

\section{A dynamical system on rigged conf\/igurations}\label{sec:3}
\subsection{Extended rigged conf\/igurations}\label{subsec:3_1}

The \textit{rigged configuration}
is an ingenious device utilized in combinatorial Bethe ansatz \cite{KKR,KR}.
We construct an example of the direct product model
associated with the rigged conf\/igurations.
The result of this section will be used in Subsection~\ref{subsec:4_4}.

We introduce a pair of sets $X_1$ and $X_2$.
Choose an arbitrary positive integer $N$ and def\/ine $X_1 = \Z^N$.
For any pair of positive integers $m$, $p$ we def\/ine
\begin{equation}\label{eq:may1_1}
\Lambda (m,p) = \{(\lambda_i)_{i \in \Z} \mid
\lambda_i \in \Z, \; \lambda_1 = 0, \; \lambda_i \le \lambda_{i+1}, \; \lambda_{i+m} = \lambda_i + p\;
\hbox{ for all } i \}.
\end{equation}
Given any positive integer sequences $(m_j)_{1 \leq j \leq N}$, $(p_j)_{1 \leq j \leq N}$
we def\/ine $X_2 = \prod_{j=1}^N \Lambda (m_j,p_j)$.
Each element of $X_2$ is written, for instance, as
$\bol = (\lambda^{(j)})_{1 \leq j \leq N}$ with $\lambda^{(j)} \in \Lambda (m_j,p_j)$
or $\bol = (\lambda^{(j)}_i)_{i \in \Z, 1 \leq j \leq N}$.

\begin{example}\label{ex:oct13_1}
Let $(m_1, m_2, m_3) = (3,2,1)$ and $(p_1, p_2, p_3) = (12,6,4)$.
Each element of the set $X_2 = \Lambda (3,12) \times \Lambda (2,6) \times \Lambda (1,4)$
is written as $\bol = (\lambda^{(1)}, \lambda^{(2)},\lambda^{(3)})$.
It is labeled by {three} integers $a=\lambda^{(1)}_2$, $b=\lambda^{(1)}_3$
and $c=\lambda^{(2)}_2$ satisfying the conditions $0 \leq a \leq b \leq 12$ and $0 \leq c \leq 6$.
Note that $\lambda^{(3)} = (\lambda^{(3)}_i)_{i \in \Z}$ has {a unique} element $\lambda^{(3)}_i = 4 (i-1)$.
\end{example}
%%%%
The set $\tilde{X} = X_1 \times X_2$ with the above $X_1$, $X_2$ is regarded as the non-compact level set
considered in Subsection~\ref{subsec:2_3}.
It will be identif\/ied with the set of {\em extended rigged configurations} in Subsection~\ref{subsec:4_3}
under a certain condition imposed on the values of $m_j$ and $p_j$.
For the time being we ignore this condition and call $\tilde{X}$ itself the set of extended rigged conf\/igurations.

We introduce a pair of abelian groups $H_1$ and $H_2$.
Suppose $H_1$ acts on $X_1$
by
\begin{displaymath}
\varGamma : \ H_1 \times X_1 \rightarrow X_1, \qquad
(g , \bom) \mapsto \varGamma (g , \bom) = \bom + \psi_{\varGamma} (g),
\end{displaymath}
where $\psi_{\varGamma}: H_1 \rightarrow X_1$ is a map common to every $\bom \in X_1$.
This map must be linear, i.e.~for all $g, h \in H_1$ we have $\psi_{\varGamma}(g+h) = \psi_{\varGamma}(g) + \psi_{\varGamma}(h)$.
We assume that this action is transitive, i.e.~for any $\bom, \bom' \in X_1$ there exists $g \in H_1$ such that $\varGamma (g , \bom) = \bom'$.
An example of such a map will appear in the next section, just above Lemma~\ref{lem:oct18_2}.
\begin{remark}
Since $X_1 = \Z^N$, any sub-lattice of $X_1$ can be thought of as an abelian group which
acts on $X_1$ by translation.
As a sub-lattice of $X_1$ we can take $X_1$ itself, which leads to the above~$H_1$ and its action on $X_1$.
\end{remark}

Suppose $H_2$ is a free abelian group with generators $s_1, \ldots , s_N$.
We def\/ine $\Upsilon: H_2 \times X_2 \rightarrow X_2$ to be an action of~$H_2$ on $X_2$ by
\begin{equation}\label{eq:mar24_1}
\Upsilon (g,\bol) = \big(\lambda^{(j)}_{n_j + i}- \lambda^{(j)}_{n_j + 1}\big)_{i \in \Z, 1 \leq j \leq N},
\end{equation}
for each $g = \sum_{j=1}^N n_j s_j\in H_2$ and
$\bol = (\lambda^{(j)}_i)_{i \in \Z, 1 \leq j \leq N} \in X_2$.

\begin{example}\label{ex:oct15_1}
Consider the $X_2$ in Example~\ref{ex:oct13_1}.
Let $\overline{\bol} = \Upsilon (g,\bol)$
for $g = n_1 s_1 + n_2 s_2 + n_3 s_3 \in H_2$.
Then
$\overline{\bol} = \big(\overline{\lambda}^{(1)}, \overline{\lambda}^{(2)},\overline{\lambda}^{(3)}\big)$ is specif\/ied by three integers
$\overline{a}=\overline{\lambda}^{(1)}_2, \overline{b}=\overline{\lambda}^{(1)}_3$
and $\overline{c}=\overline{\lambda}^{(2)}_2$ as
\begin{equation*}
(\overline{a}, \overline{b}) =
\begin{cases}
(a,b) & n_1 \equiv 0 \pmod{3},\\
(b-a, 12-a) & n_1 \equiv 1 \pmod{3},\\
(12-b, 12-b+a) & n_1 \equiv 2 \pmod{3},
\end{cases}
\qquad
\overline{c} =
\begin{cases}
c & n_2 \equiv 0 \pmod{2},\\
6-c & n_2 \equiv 1 \pmod{2}.
\end{cases}
\end{equation*}
Here $a$, $b$, $c$ are the labels for $\bol$ in Example \ref{ex:oct13_1}.
\end{example}

\subsection{Cyclic group structures}\label{subsec:3_2}

Let $\rho_{\Upsilon}: H_2 \rightarrow {\frak S}(X_2)$ be
the permutation representation for the
action of $\Upsilon$ in the previous subsection.
Then
$\rho_{\Upsilon} (H_2)$ is isomorphic to $\mathcal{C}_{m_1} \times \cdots \times \mathcal{C}_{m_N}$, where
$\mathcal{C}_m$ is an order $m$ cyclic group.
This in particular determines an action of $\mathcal{C}_m$ on $\Lambda (m,p)$.
The following facts are well known in the theory of cyclic groups.
\begin{proposition}
Every subgroup of a cyclic group is a cyclic group.
\end{proposition}
\begin{proposition}
Let $m$ be a positive integer and $n$ be a divisor of $m$.
Then  there exists a unique subgroup of $\mathcal{C}_m$ that is isomorphic to $\mathcal{C}_n$.
\end{proposition}
Let $\alpha$ be a common divisor of $m$ and $p$.
As a subset of $\Lambda (m,p)$ % in (\ref{eq:may1_1})
we def\/ine
\begin{gather}\label{eq:may12_1}
\Lambda^{(\alpha)} (m,p) = \{
\lambda \in \Lambda (m/\alpha,p/\alpha) \mid
\lambda \notin \Lambda (m/\alpha',p/\alpha') \, \mbox{for every common divisor} \, \alpha' > \alpha  \}.\!\!
\end{gather}
In other words
$\Lambda^{(\alpha)} (m,p)$ is the set of all elements of $\Lambda (m,p)$
whose stabilizer associated with
the action of {$\mathcal{C}_m$}
is isomorphic to $\mathcal{C}_\alpha$.
Now we have the decomposition $\Lambda (m,p) = \bigsqcup_{\alpha} \Lambda^{(\alpha)}(m,p)$
where $\alpha$ runs over every common divisor of $m$ and $p$.

Choose a sequence $\boa = (\alpha_j)_{1 \leq j \leq N}$ with
each $\alpha_j$ being a common divisor of $m_j$ and $p_j$.
Let $H_2^{(\boa)}$ be a group generated by $(m_1/\alpha_1) s_1, \ldots ,(m_N/\alpha_N) s_N$.
It is a subgroup of $H_2$ whose
image $\rho_{\Upsilon} (H_2^{(\boa)})$ by the permutation representation
% is a subgroup of
% $\rho_{\Upsilon} (H_2) \cong \mathcal{C}_{m_1} \times \cdots \times \mathcal{C}_{m_N}$ that
is isomorphic to $\mathcal{C}_{\alpha_1} \times \cdots \times \mathcal{C}_{\alpha_N}$.
Let
$X_2^{(\boa)} = \prod_{j=1}^N \Lambda^{(\alpha_j)} (m_j,p_j)$.
In other words $X_2^{(\boa)}$ is the set of all elements of $X_2$ whose stabilizer
associated with the $H_2$ action is $H_2^{(\boa)}$.
\begin{example}\label{ex:oct15_2}
Consider the $X_2$ in Example \ref{ex:oct13_1}.
We have $\Lambda (3,12) = \Lambda^{(1)}(3,12) \sqcup \Lambda^{(3)}(3,12)$,
$\Lambda (2,6) = \Lambda^{(1)}(2,6) \sqcup \Lambda^{(2)}(2,6)$ and
$\Lambda (1,4) = \Lambda^{(1)}(1,4)$.
If we set $\boa = (\alpha_1, \alpha_2, \alpha_3) = (3,1,1)$,
then
$(a,b) = (4,8)$ and $c \ne 3$ for any $\bol \in X_2^{(\boa)}$.
It follows from Example \ref{ex:oct15_1} that
the set of all $g \in H_2$ satisfying $\Upsilon (g, \bol) = \bol$
is
$H_2^{(\boa)} = \{ g=  n_1 s_1 + 2 n_2 s_2 + n_3 s_3\, | \,n_1, n_2, n_3 \in \Z \}$.
\end{example}

Since $H_2$ is abelian, $X_2^{(\boa)} (\subset X_2)$ is invariant under the action of $H_2$.
Let
\begin{equation}\label{eq:apr22_2}
X^{(\boa)} = X_1 \times X_2^{(\boa)}.
\end{equation}

The $X_2^{(\boa)}$ here corresponds to the $X_2'$, and the $X^{(\boa)}$ to the $X$ in Subsection~\ref{subsec:2_2}.
The set
$\tilde{X} = X_1 \times X_2$ is decomposed as
\begin{math}
\tilde{X} = \bigsqcup_{\boa} X^{(\boa)}
\end{math}
where each $\alpha_j$ runs over every common divisor of $m_j$ and $p_j$.

\subsection{Group actions on the set of rigged conf\/igurations}\label{subsec:3_3}
We def\/ine a pair of commutative actions of $H_1$ and $H_2$ on $\tilde{X} = X_1 \times X_2$.
As in Subsection \ref{subsec:2_2}, let $\varGamma_1 : H_1 \times \tilde{X} \rightarrow \tilde{X}$
be a diagonal action of
$H_1$ on $\tilde{X}$
in which its action on $X_2$ is trivial, i.e.~for each $f \in H_1$ and $x =(\bom,\bol) \in \tilde{X}$
we set
$\varGamma_1(f, x) = (\bom + \psi_\varGamma (f), \bol)$.
%%%
Let
$\varGamma_2 : H_2 \times \tilde{X} \rightarrow \tilde{X}$
% be an action of $H_2$ on $\tilde{X}$
be an action of
$H_2$ on $\tilde{X}$
which is def\/ined as follows.

Fix an arbitrary set of $N^2$ integers $(B_{j,k})_{1 \leq j,k \leq N} \in \Z^{N^2}$.
In the next section we take $B_{j,k}$ to be those given in (\ref{eq:may26_1}).
For each $g \in H_2$ and $x =(\bom,\bol) \in \tilde{X}$
we set
$\varGamma_2(g, x) = (\varphi (g, \bom, \bol), \Upsilon (g,\bol))$ where
{$\Upsilon (g,\bol)$ is given by (\ref{eq:mar24_1}) and
$\varphi : H_2 \times X_1 \times X_2 \rightarrow X_1$ is a map def\/ined by
\begin{equation}\label{eq:oct19_2}
\varphi (g, \bom, \bol) = \bom +
% (\phi_j(g, \lambda^{(j)}))_{1 \leq j \leq N},
\left(  \lambda^{(j)}_{n_j+1} + \sum_{k=1}^N B_{j,k} n_k \right)_{1 \leq j \leq N},
\end{equation}
for each $g = \sum_{k=1}^N n_k s_k \in H_2$,
$\bom \in X_1$ and
$\bol = (\lambda^{(j)}_i)_{i \in \Z, 1 \leq j \leq N} \in X_2$.}
It is easy to see that
the map $\varGamma_2$ indeed yields an action of $H_2$ on $\tilde{X}$, since
the relation $\varphi (g + h, \bom, \bol) =
\varphi (h, \varphi(g, \bom, \bol) , \Upsilon (g,\bol))$ holds for all $g,h \in H_2$.
It is also easy to see that
$H_1$ and $H_2$ act on $\tilde{X}$ commutatively by $\varGamma_1$ and $\varGamma_2$.
\begin{example}\label{ex:oct15_3}
Consider the $X_2$ in Example \ref{ex:oct13_1}.
For $g = n_1 s_1 + n_2 s_2 + n_3 s_3 \in H_2$ and
$x =(\bom,\bol) \in \tilde{X}$
we write $\varGamma_2(g, x) =(\overline{\bom}, \overline{\bol}) \in \tilde{X}$.
Here $\overline{\bol} = \Upsilon (g , \bol)$ is the one given in Example \ref{ex:oct15_1}, and
$\overline{\bom} = \varphi (g, \bom, \bol)$ is written as
\begin{displaymath}
\varphi (g, \bom, \bol) = \bom +
\left( \mu_j + \sum_{k=1}^3 B_{j,k} n_k \right) _{1 \leq j \leq 3}.
\end{displaymath}
Here we set
\begin{equation*}
\mu_1 =
\begin{cases}
12 l_1 & n_1 = 3 l_1,\\
12 l_1 + a & n_1 = 3 l_1+1,\\
12 l_1 + b & n_1 = 3 l_1+2,
\end{cases}
\qquad
\mu_2 =
\begin{cases}
6 l_2 & n_2 = 2 l_2,\\
6 l_2 + c & n_2 = 2 l_2 + 1,
\end{cases}
\end{equation*}
for any $l_1, l_2 \in \Z$, and $\mu_3 = 4 n_3$.
The $a$, $b$, $c$ are the labels for $\bol$ in Example~\ref{ex:oct13_1}.
\end{example}

By def\/inition the set $X^{(\boa)} (\subset \tilde{X})$ is invariant under the action of $H_1$.
It is also invariant under~$H_2$.
Thus by restricting their domains from $\tilde{X}$ to $X^{(\boa)}$, $\varGamma_1$ and $\varGamma_2$
yield their actions on~$X^{(\boa)}$.
In what follows we denote them by the same symbols $\varGamma_1$ and $\varGamma_2$
for simplicity.

We def\/ine
$G_1 = \rho_{\varGamma_1} (H_1), G_2 = \rho_{\varGamma_2} (H_2)$, and
$G_2^{(\boa)} = \rho_{\varGamma_2} (H_2^{(\boa)})$
to be subgroups of ${\frak S}(X^{(\boa)})$
where $\rho_{\varGamma_1}$ and $\rho_{\varGamma_2}$ are
the permutation representations.

\begin{lemma}\label{lem:oct19_1}
Consider the action of~$G_2^{(\boa)}$ on $X^{(\boa)}= X_1 \times X_2^{(\boa)}$.
\begin{enumerate}\itemsep=0pt
\item It is trivial on the $X_2^{(\boa)}$-part.
\item On the $X_1$-part, $G_2^{(\boa)}$ acts as a sub-lattice of~$X_1$ by translation.
\item The action on the $X_1$-part is independent of the $X_2^{(\boa)}$-part.
\end{enumerate}
\end{lemma}

\begin{proof}
Since the $G_2^{(\boa)}$ corresponds to the $G_2'$ in Subsection \ref{subsec:2_2}, we have
$G_2^{(\boa)} = G_1 \cap G_2$ by Lemma~\ref{lem:feb9_2}.
In particular $G_2^{(\boa)}$ is a subgroup of $G_1$, hence follow items $(i)$ and $(ii)$.
Consider item~$(iii)$.
The $X_1$-part is given by the map $\varphi$ in (\ref{eq:oct19_2}).
We show that
for each $g = \sum_{k=1}^N n_k (m_k/\alpha_k) s_k \in H_2^{(\boa)}$
and $\bom \in X_1$ the image $\varphi(g,\bom,\bol)$ of the map $\varphi$
is common to all $\bol \in X_2^{(\boa)}$.
In fact for any $\bol \in X_2$
we have
\begin{equation}\label{eq:may1_2}
\varphi(g,\bom,\bol) = \bom + \left(
\lambda^{(j)}_{(m_j/\alpha_j) n_j+1} + \sum_{k=1}^N (m_k/\alpha_k) B_{j,k} n_k
\right)_{1 \leq j \leq N}.
\end{equation}
Note that if $\bol$ lies in $X_2^{(\boa)}$ then we have $\lambda^{(j)}_{(m_j/\alpha_j) n_j+1} = (p_j/\alpha_j) n_j$ from
(\ref{eq:may1_1}) and (\ref{eq:may12_1}).
Hence
$\varphi(g,\bom,\bol)$ is independent of the choice of $\bol \in X_2^{(\boa)}$.
\end{proof}

We def\/ine $F^{(\boa)} = (F^{(\boa)}_{j,k})_{1 \leq j,k \leq N}$ to be an $N \times N$ matrix whose elements are given by
\begin{equation}\label{eq:may15_1}
F^{(\boa)}_{j,k} = (\delta_{j,k} p_k + B_{j,k} m_k)/\alpha_k ,
\end{equation}
and regard $\bom$, $\boldsymbol{n} = {}^t (n_1,\ldots,n_N)$ and $\varphi(g,\bom,\bol)$ as column vectors.
Then (\ref{eq:may1_2}) is written as
$\varphi(g,\bom,\bol) = \bom + F^{(\boa)} \boldsymbol{n}$
when $\bol$ lies in $X_2^{(\boa)}$.
By Lemma~\ref{lem:oct19_1} we can
think of $G_2^{(\boa)}$ as a
subgroup of ${\frak S}(X_1)$ rather than that of ${\frak S}(X^{(\boa)})$.
In this sense the group $G_2^{(\boa)}$ is isomorphic to the lattice $F^{(\boa)} \Z^N$
which acts on $X_1 = \Z^N$ by translation.

Thus Theorem~\ref{th:apr10_1} for the present construction with Lemma~\ref{lem:apr8_2}
(to appear in the appendix)
is now stated as follows.
\begin{proposition}\label{prop:apr22_1}
Let the set $X^{(\boa)}$, the matrix $F^{(\boa)}$, the groups $H_1$, $H_2$ and their actions on $X^{(\boa)}$
be def\/ined as above, and
$x$ be any element of $X^{(\boa)}$.
Then there is a bijection between $H_1 \cdot [x]_{H_2}$ and $F^{(\boa)} \Z^N \backslash \Z^N$
that is commutative with the action of $H_1$.
\end{proposition}
The set $F^{(\boa)} \Z^N \backslash \Z^N$ becomes compact {(and is a torus)} if and only if $\det F^{(\boa)} \ne 0$.
A way to achieve this condition is given as follows.
Let $L$ be an integer satisfying $L \geq 2 \sum_{k=1}^N k m_k$ and set
\begin{gather}\label{eq:may15_2}
p_j = L - 2\sum_{k =1}^N \min(j,k)m_k,\\
B_{j,k} = 2 \min (j,k).\label{eq:may26_1}
\end{gather}
Then we have $L > p_1 > p_2 > \cdots > p_N = L - 2 \sum_{k=1}^N k m_k \geq 0$ and
\begin{equation*}
\det F^{(\boa)} = L p_1 p_2 \cdots p_{N-1}/(\alpha_1 \cdots \alpha_N) > 0.
\end{equation*}
Note that $p_N=0$ is allowed here.
In this case we have $\Lambda (m_N,p_N) = \Lambda^{(m_N)} (m_N,p_N)$ which has {a unique} element $(0)_{i \in \Z}$.

\begin{example}\label{ex:oct15_4}
The $p_j$ and $m_j$ in Example \ref{ex:oct13_1} satisfy the relation (\ref{eq:may15_2}) with $L=24$.
We set $\boa = (\alpha_1, \alpha_2, \alpha_3) = (3,1,1)$ as in Example \ref{ex:oct15_2}.
By taking $B_{j,k}$ as in (\ref{eq:may26_1}) we have
\begin{gather*}
F^{(\boa)}  =
\begin{pmatrix}
p_1/\alpha_1  & 0 & 0 \\
0 &  p_2/\alpha_2  & 0 \\
0 & 0 &   p_3/\alpha_3
\end{pmatrix}
+
\begin{pmatrix}
2 & 2 & 2 \\
2 & 4 & 4 \\
2 & 4 & 6
\end{pmatrix}
\begin{pmatrix}
m_1/\alpha_1   & 0 & 0 \\
0 &  m_2/\alpha_2  & 0 \\
0 & 0 &   m_3/\alpha_3
\end{pmatrix}\\
\phantom{F^{(\boa)}}{}
 =
\begin{pmatrix}
6 & 4 & 2 \\
2 & 14 & 4 \\
2 & 8 & 10
\end{pmatrix}
,
\end{gather*}
and $\det F^{(\boa)} = 576$.
For instance, let $x = (\bom, \bol)$ be the element of $X^{(\boa)}$ specif\/ied by
$\bom = (0,0,0) \in X_1 = \Z^3$, and
$\bol \in X_2^{(\boa)}$ that is labeled by $(a,b,c) = (4,8,1)$.
Then there is a bijection between $H_1 \cdot [x]_{H_2}$ and
the three dimensional torus $F^{(\boa)} \Z^3 \backslash \Z^3$ that is commutative with the action of $H_1$.
\end{example}

%%%%%%%%%%%%%%%%%%%%%%%%%
%
\begin{remark}\label{rem:mar12_1}
Given any positive integer $s$, choose an increasing sequence of positive integers $j_1, \ldots ,j_s$
and def\/ine $\H$ to be the set $\H=\{j_1,\ldots,j_s\}$.
{A generalization of the arguments in this section is given by} replacing the positive integer sequence $(m_j)_{1 \leq j \leq N}$ by
$(m_j)_{j \in \H}$, which f\/its to the case in \cite{KTT}.

All the results in this paper can be extended to this generalized setting:
We set $X_1 = \Z^s$, and
$H_2$ to be a free abelian group with generators $\{ s_j \}_{j \in \H}$.
Given any {positive} integer sequence $(p_j)_{j \in \H}$
we set $X_2 = \prod_{j \in \H} \Lambda (m_j,p_j)$.
Proposition \ref{prop:apr22_1} becomes a claim on the bijection between
$H_1 \cdot [x]_{H_2}$ and $F^{(\boa)} \Z^s \backslash \Z^s$ where
the $F^{(\boa)}= (F^{(\boa)}_{j,k})_{j,k \in \H}$ is now an $s \times s$ matrix.
The other def\/initions and statements can be modif\/ied similarly.

\end{remark}
%%%%%%%%%%%%%%%%%%%%%%%%%%%%%%%%%%%%%%%%%%%%%%%%%%%%%%%%%%%%%%%%
\section{A one-dimensional integrable cellular automaton}\label{sec:4}
\subsection{The periodic box-ball system}\label{subsec:4_1}

The periodic box-ball system (pBBS) is a one-dimensional cellular automaton with periodic boundary conditions.
We give a brief review on this system based on~\cite{KTT} from here to Subsection~\ref{subsec:4_3}.
Let $L$ be a positive integer and $p$ be a sequence of letters $1$ and $2$ under the conditions
$\# (1) \geq \# (2)$ and $\# (1) + \# (2) = L$.
Such sequences are called {\em paths} of positive weight and of length $L$.
Denote by ${\mathcal P}$ the set of all such paths.
We can def\/ine a commuting family of time evolutions $T_k$ $(k=1,2,\ldots)$ acting on ${\mathcal P}$.
It is a collection of update procedures for the cellular automaton.
In this paper we write $n T_k$ for an $n$ times repeated application of $T_k$ instead of $T_k^n$,
regarding $T_k$s as generators of an abelian group (see Subsection~\ref{subsec:4_3}).

The action of $T_1$ is given by a cyclic shift by one digit to the right.
The def\/inition of $T_k$ for the other $k$'s is given by means of the crystal basis of
the quantized envelope algebra \cite{K}
and is available in Section~2.2 of~\cite{KTT}.
Here we review it shortly.

Let $B_k$ be the set of one-row semistandard tableaux of length $k$ with entries $1$ and $2$.
For instance, $B_1 = \{ 1,2 \}$, $B_2 = \{ 11, 12, 22 \}$ and $B_3 = \{ 111, 112, 122, 222 \}$.
The combinatorial $R$ map $R: B_k \times B_1 \rightarrow B_1 \times B_k$ is def\/ined as follows.
If we depict the relation $R(x,y) = (\tilde{y}, \tilde{x})$ by
\begin{equation*}
\begin{picture}(90,40)(-20,-9)
\unitlength 0.4mm
\put(0,10){\line(1,0){20}}\put(-6,8){$x$}\put(22,8){${\tilde x}$}
\put(10,0){\line(0,1){20}}\put(8,25){$y$}\put(8,-9){${\tilde y}$}
%
% \put(47,8){or}
% \put(80,10){\line(1,0){20}}\put(74,8){${\tilde y}$}\put(102,8){$y$}
% \put(90,0){\line(0,1){20}}\put(88,25){${\tilde x}$}\put(88,-9){$x$}
\end{picture}
\end{equation*}
then the def\/inition of $R$ is given by
the following diagrams:
%\begin{figure}[!th]%\label{fig:cr}
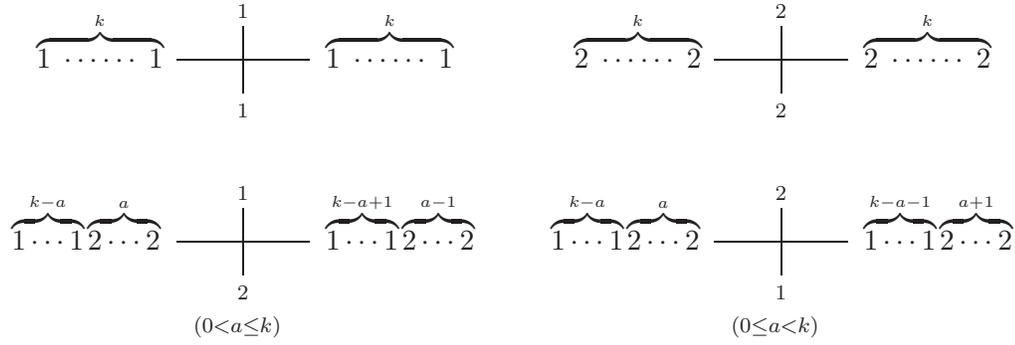
\begin{center}\hspace*{-85mm}
% \begin{picture}(130,120)(55,0)
\begin{picture}(130,120)(-60,15)
\unitlength 1.1mm
\multiput(0,3)(0,22){2}{
\multiput(2,10)(65,0){2}{\line(1,0){16}}
\multiput(10,6)(65,0){2}{\line(0,1){8}}}

\put(9.3,40){$\scriptstyle 1$}
\put(-15,34){$\scriptstyle \overbrace{1 \;\cdots \cdots \; 1}^k$}
\put(20,34){$\scriptstyle \overbrace{1 \;\cdots \cdots \; 1}^k$}
\put(9.3,28){$\scriptstyle 1$}
% \put(7,24){$\scriptstyle H=0$}

\put(74.3,40){$\scriptstyle 2$}
\put(50,34){$\scriptstyle \overbrace{2 \;\cdots \cdots \; 2}^k$}
\put(85,34){$\scriptstyle \overbrace{2 \;\cdots \cdots \; 2}^k$}
\put(74.3,28){$\scriptstyle 2$}
% \put(72,24){$\scriptstyle H=0$}

\put(9.3,18){$\scriptstyle 1$}
\put(-18,12){$\scriptstyle \overbrace{1 \cdots 1}^{k-a}
\overbrace{2\cdots 2}^a$}
\put(20,12){$\scriptstyle \overbrace{1 \cdots 1}^{k-a+1}
\overbrace{2\cdots 2}^{a-1}$}
\put(9.3,6){$\scriptstyle 2$}
\put(4,2){$\scriptstyle (0<a\le k)$}

\put(74.3,18){$\scriptstyle 2$}
\put(47.2,12){$\scriptstyle \overbrace{1 \cdots 1}^{k-a}
\overbrace{2\cdots 2}^a$}
\put(85,12){$\scriptstyle \overbrace{1 \cdots 1}^{k-a-1}
\overbrace{2\cdots 2}^{a+1}$}
\put(74.3,6){$\scriptstyle 1$}
\put(69,2){$\scriptstyle (0\le a<k)$}
\end{picture}
% \caption{Combinatorial $R: B_l \otimes B_1 \simeq B_1 \otimes B_l$}
\vspace{1mm}
\end{center}

By repeated use of this $R$ we def\/ine the following map
\begin{gather}
B_k \times (B_1 \times \cdots \times B_1)   \rightarrow  (B_1 \times \cdots \times B_1) \times B_k, \nonumber \\
v \times (b_1 \times \cdots \times b_L)   \mapsto  (b'_1 \times \cdots \times b'_L) \times v'.\label{eq:oct9_1}
\end{gather}
\begin{example}\label{ex:oct9_2}
Set $k=3$, $L=13$, $v=122$ and $b_1 \ldots b_{13} = 1122211112212$.
(We omit the symbol~$\times$ here and in what follows.)
Then we have $b'_1 \ldots b'_{13} = 2211122211121$ and $v'=122$.
It is verif\/ied by the following diagram
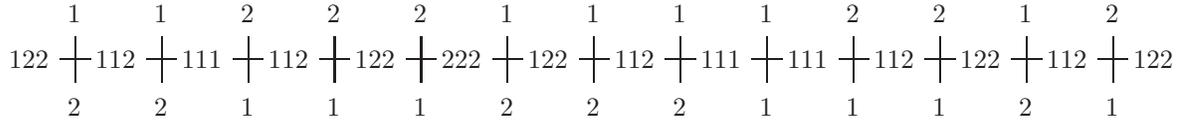
\begin{center}
\unitlength 1mm
{\small \hspace*{3mm}
\begin{picture}(130,20)(8,-11)
\multiput(0.8,0)(11.5,0){13}{\line(1,0){4}}
\multiput(2.8,-3)(11.5,0){13}{\line(0,1){6}}
\put(1.8,5){1}
\put(13.3,5){1}
\put(24.8,5){2}
\put(36.3,5){2}
\put(47.8,5){2}
\put(59.3,5){1}
\put(70.8,5){1}
\put(82.3,5){1}
\put(93.8,5){1}
\put(105.3,5){2}
\put(116.8,5){2}
\put(128.3,5){1}
\put(139.8,5){2}

\put(1.8,-7.5){2}
\put(13.3,-7.5){2}
\put(24.8,-7.5){1}
\put(36.3,-7.5){1}
\put(47.8,-7.5){1}
\put(59.3,-7.5){2}
\put(70.8,-7.5){2}
\put(82.3,-7.5){2}
\put(93.8,-7.5){1}
\put(105.3,-7.5){1}
\put(116.8,-7.5){1}
\put(128.3,-7.5){2}
\put(139.8,-7.5){1}

\put(-6,-1.1){122}
\put(5.5,-1.1){112}
\put(17,-1.1){111}
\put(28.5,-1.1){112}
\put(40,-1.1){122}
\put(51.5,-1.1){222}
\put(63,-1.1){122}
\put(74.5,-1.1){112}
\put(86,-1.1){111}
\put(97.5,-1.1){111}
\put(109,-1.1){112}
\put(120.5,-1.1){122}
\put(132,-1.1){112}
\put(143.5,-1.1){122}

\end{picture}
}
\end{center}
\end{example}

In this example we have $v = v'$.
In fact one can always f\/ind an element of $B_k$ with this property.

\begin{proposition}[\cite{KTT}]
Given any $b_1 \times \cdots \times b_L \in {\mathcal P} \subset B_1^{\times L}$,
let $v_0 \in B_k$ be the one defined in the same way as \eqref{eq:oct9_1} by
\begin{gather*}
\overbrace{1\ldots 1}^{k} \times (b_1 \times \cdots \times b_L)   \mapsto  (b''_1 \times \cdots \times b''_L) \times v_0.
\end{gather*}
Then we have $v=v'$ in \eqref{eq:oct9_1} when we adopt this $v_0$ as the $v$ there.
\end{proposition}

By this choice of $v$, we def\/ine the time evolution $T_k$ by
$T_k (b_1 \dots b_L) = b'_1 \dots  b'_L$.
\begin{example}
We have
$T_3 (1122211112212) = 2211122211121$ by Example \ref{ex:oct9_2}.
\end{example}

\begin{remark}\label{rem:may19_1}
Given any initial condition,
the actions of $T_k$s for suf\/f\/iciently large $k$'s are common to them and determine a unique time evolution.
The original study of pBBS \cite{YT, YYT} considers only this time evolution which we denote by $T_\infty$.
{In the context of the dynamical system in Subsection~\ref{subsec:2_3},
we can regard the $T_\infty$ as the unique time evolution in this dynamical system, and the other $T_k$s can be thought of as the phase f\/lows
associated with the other f\/irst integrals.}
\end{remark}

Here we give an example of the time evolution of this cellular automaton. (The case of $L=31$, and evolved by $T_4=T_\infty$.)

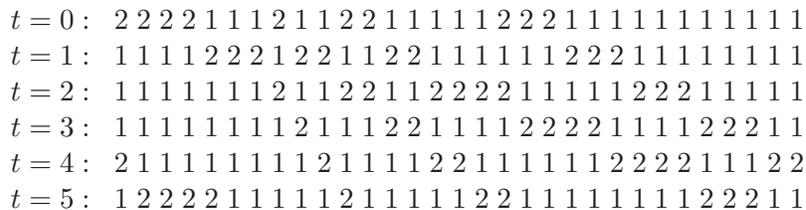
\begin{center}
$t=0: \;\;2\; 2\; 2\; 2\; 1\; 1\; 1\; 2\; 1\; 1\; 2\; 2\; 1\; 1\; 1\; 1\; 1\; 2\; 2\; 2\; 1\; 1\; 1\; 1\; 1\; 1\; 1\; 1\; 1\; 1\; 1$\\
$t=1: \;\;1\; 1\; 1\; 1\; 2\; 2\; 2\; 1\; 2\; 2\; 1\; 1\; 2\; 2\; 1\; 1\; 1\; 1\; 1\; 1\; 2\; 2\; 2\; 1\; 1\; 1\; 1\; 1\; 1\; 1\; 1$\\
$t=2: \;\;1\; 1\; 1\; 1\; 1\; 1\; 1\; 2\; 1\; 1\; 2\; 2\; 1\; 1\; 2\; 2\; 2\; 2\; 1\; 1\; 1\; 1\; 1\; 2\; 2\; 2\; 1\; 1\; 1\; 1\; 1$\\
$t=3: \;\;1\; 1\; 1\; 1\; 1\; 1\; 1\; 1\; 2\; 1\; 1\; 1\; 2\; 2\; 1\; 1\; 1\; 1\; 2\; 2\; 2\; 2\; 1\; 1\; 1\; 1\; 2\; 2\; 2\; 1\; 1$\\
$t=4: \;\;2\; 1\; 1\; 1\; 1\; 1\; 1\; 1\; 1\; 2\; 1\; 1\; 1\; 1\; 2\; 2\; 1\; 1\; 1\; 1\; 1\; 1\; 2\; 2\; 2\; 2\; 1\; 1\; 1\; 2\; 2$\\
$t=5: \;\;1\; 2\; 2\; 2\; 2\; 1\; 1\; 1\; 1\; 1\; 2\; 1\; 1\; 1\; 1\; 1\; 2\; 2\; 1\; 1\; 1\; 1\; 1\; 1\; 1\; 1\; 2\; 2\; 2\; 1\; 1$\\
$t=6: \;\;2\; 1\; 1\; 1\; 1\; 2\; 2\; 2\; 2\; 1\; 1\; 2\; 1\; 1\; 1\; 1\; 1\; 1\; 2\; 2\; 1\; 1\; 1\; 1\; 1\; 1\; 1\; 1\; 1\; 2\; 2$\\
$t=7: \;\;1\; 2\; 2\; 2\; 1\; 1\; 1\; 1\; 1\; 2\; 2\; 1\; 2\; 2\; 2\; 1\; 1\; 1\; 1\; 1\; 2\; 2\; 1\; 1\; 1\; 1\; 1\; 1\; 1\; 1\; 1$\\
$t=8: \;\;1\; 1\; 1\; 1\; 2\; 2\; 2\; 1\; 1\; 1\; 1\; 2\; 1\; 1\; 1\; 2\; 2\; 2\; 2\; 1\; 1\; 1\; 2\; 2\; 1\; 1\; 1\; 1\; 1\; 1\; 1$\\
$t=9: \;\;1\; 1\; 1\; 1\; 1\; 1\; 1\; 2\; 2\; 2\; 1\; 1\; 2\; 1\; 1\; 1\; 1\; 1\; 1\; 2\; 2\; 2\; 1\; 1\; 2\; 2\; 2\; 1\; 1\; 1\; 1$
% $t=10: \;\;1\; 1\; 1\; 1\; 1\; 1\; 1\; 1\; 1\; 1\; 2\; 2\; 1\; 2\; 2\; 1\; 1\; 1\; 1\; 1\; 1\; 1\; 2\; 2\; 1\; 1\; 1\; 2\; 2\; 2\; 2$
\end{center}

If we denote by $p \in {\mathcal P}$ the sequence for $t=0$ then the sequence for $t=n$ stands for
$(n T_4) \cdot p \in {\mathcal P}$.

When suf\/f\/iciently separated from the other $2$'s, one can think of
a consecutive sequence of $2$'s of length $k$ as a soliton of amplitude $k$.
By an appropriate def\/inition one can say that the number of solitons conserves for each amplitude.
In this example there are four solitons of distinct amplitudes (1, 2, 3 and 4) in every time step.
For instance, in the state for $t=9$ in the above example, the sequence $2\; 2\; 2\; 1\; 1\; 2\; 2\; 2$ should be interpreted as
an intermediate stage of the collision of two solitons of amplitudes $4$ and $2$.

\subsection{Soliton content and rigged conf\/igurations}\label{subsec:4_2}

Let
$\Pth_+$ be the set of all ballot sequences in $\Pth $: A path $p = b_1 b_2 \ldots b_L$ is called a ballot sequence if and only if
the number of $1$'s in the pref\/ix $b_1 \ldots b_k$ is at least as large as the number of $2$'s in that pref\/ix
for every $k$.
For any $p \in {\mathcal P}$ there exist an integer $d$ and a sequence $p_+ \in \Pth_+$ such that $p = (d T_1) \cdot p_+$
where $T_1$ is the right cyclic shift operator.

Given any $p_+ \in \Pth_+$, a bijective map $\phi$
due to Kerov--Kirillov--Reshetikhin \cite{KKR,KR}
yields the following data
$(\H, \m, \boJ)$.
Here $\H=\{ j_1,  \ldots ,j_s \}$ is a set of positive integers
satisfying the condition $j_1 < \dots < j_s$, and
$\m=( m_{j_1},\ldots,m_{j_s} )$ is an array of positive integers.
The
$\boJ = \left((J^{(j_1)}_i)_{1 \le i \le m_{j_1}}, \ldots,
(J^{(j_s)}_i)_{1 \le i \le m_{j_s}}\right)$
is an array of partitions:
Each $(J^{(j)}_i)_{1 \le i \le m_{j}}$ is a partition of non-negative integers into at most $m_j$ parts
in increasing order,
with largest part $\le p_j$
where
\begin{math}% \label{eq:va}
p_j = L - 2\sum_{k \in \H}\min(j,k)m_k.
\end{math}
In symbols it is written as $\phi (p_+) = (\m,\boJ)$.
The collection $(\m,\boJ)$ is called a rigged conf\/iguration with conf\/iguration $\m$ and
riggings $\boJ$, and $p_j$ are called \textit{vacancy numbers}.
The set of all rigged conf\/igurations with conf\/iguration $\m$ is denoted $\Rig(\m)$.
An algorithm to obtain the data
$(\H, \m, \boJ)$ is presented in Appendix~\ref{app:B}.

Suppose there are two possible ways to write $p$ as $p =(d T_1) \cdot p_+ = (d' T_1) \cdot p'_+$.
Then it can be shown that
one obtains a common conf\/iguration $\m$ by $\phi (p_+) = (\m,\boJ)$ and $\phi (p'_+) = (\m,\boJ')$ with some $\boJ$, $\boJ'$.
Hence for each $p \in {\mathcal P}$ one can label it with a specif\/ic $\m$, and we say that such a path has {\em soliton content} $\m$.
The set ${\mathcal P}$ is decomposed as ${\mathcal P} = \bigsqcup_{\m} {\mathcal P}(\m)$
where ${\mathcal P}(\m)$ is the set of all paths of soliton content $\m$.
It can be shown that each ${\mathcal P}(\m)$ is
invariant under the action of $T_k$ for all $k$.

\begin{example}\label{ex:oct16_1}
Consider the paths
\begin{equation*}
p=p_+=121122111212211222121111 \qquad \mbox{and} \qquad p'_+=111212211222121111121122,
\end{equation*}
in $\Pth_+$ with $L=24$ which are related by
$p = (6 T_1) \cdot p'_+$.
By the bijective map $\phi$
we obtain
% \footnote{A description of the combinatorial algorithm to realize this map $\phi$ is available in Appendix A of \cite{KTT}.}
$\phi (p) = (\m, \boJ)$ and $\phi (p'_+) = (\m, \boJ')$
with
\begin{gather*}
\m  = (m_1, m_2, m_3) = (3,2,1),\\
\boJ  = \big(\big(J^{(1)}_1,J^{(1)}_2,J^{(1)}_3\big),\big(J^{(2)}_1,J^{(2)}_2\big),\big(J^{(3)}_1\big)\big) = ((0,4,8),(0,1),(0)),\\
\boJ'  = \big(\big(J'^{(1)}_1,J'^{(1)}_2,J'^{(1)}_3\big),\big(J'^{(2)}_1,J'^{(2)}_2\big),\big(J'^{(3)}_1\big)\big) = ((2,6,10),(1,6),(0)).
\end{gather*}
\end{example}

\subsection{Direct scattering transform}\label{subsec:4_3}

By means of the map $\phi$ one can construct a bijection between the set of
all states in the cellular automaton and the set of extended rigged conf\/igurations divided by a group action.
The map from the former to the latter is called the direct scattering transform
(and its inverse is referred to as the inverse scattering transform),
named after a similar transform in the theory of integrable non-linear evolution equations \cite{AS}.

For the sake of simplicity we assume $\H=\{1,\ldots,N\}$ in what follows.
Then we have $\m = (m_j)_{1 \leq j \leq N}$, $\boJ=((J^{(j)}_i)_{1 \le i \le m_{j}})_{1 \leq j \leq N}$
and the vacancy numbers are given by (\ref{eq:may15_2}).
See Remark~\ref{rem:mar12_1} for the recipe to recover the original setting for general $\H$.

We def\/ine
\begin{equation}
{\Jb} (m,p) = \{(J_i)_{i \in \Z} \mid
J_i \in \Z, \; J_i \le J_{i+1}, \; J_{i+m} = J_i + p\;
\hbox{ for all } i \},\label{eq:jcomp}
\end{equation}
and $\Jb (\m) = \prod_{j=1}^N {\Jb} (m_j,p_j)$.
There is a map
\begin{equation}\label{eq:iota}
\begin{split}
\iota : \quad
\Rig(\m) \qquad \quad\;\;
\qquad&\longrightarrow \qquad\quad
\Jb(\m)\\
((J^{(j)}_i)_{1 \le i \le m_{j}})_{1 \leq j \leq N}
&\mapsto
{(J^{(j)}_i)_{i \in \Z, 1 \leq j \leq N}}
\end{split}
\end{equation}
where the inf\/inite sequence $({J}^{(j)}_i)_{i \in \Z} \in {\Jb} (m_j,p_j)$
is the one that extends
$(J^{(j)}_i)_{1 \le i \le m_j}$ quasi-periodically as
${J}^{(j)}_{i+m_j}={J}^{(j)}_i+p_j$ for all $i \in \Z$.
{In this article we call the elements of $\Jb (\m)$ extended rigged conf\/igurations}.
{The relation between $\Jb (\m)$ and the set $\tilde{X}$ in Subsection \ref{subsec:3_1} will be explained in the next subsection.}

Let ${\mathcal A}$ be a free abelian group with generators $\sigma_1,\ldots,\sigma_N$.
We def\/ine its action
on the set~$\Jb (\m)$.
Given any $g = \sum_{k=1}^N n_k \sigma_k \in {\mathcal A}$ and
$\boJ = (J_i^{(j)})_{i \in \Z, 1 \leq j \leq N} \in \Jb (\m)$
we set
\begin{displaymath}
g \cdot \boJ = \left( J_{i+n_j}^{(j)} + 2 \sum_{k=1}^N n_k \min (j,k) \right)_{i \in \Z, 1 \leq j \leq N}.
\end{displaymath}

Recall the family of time evolutions $T_k$ $(k=1,2,\ldots)$ introduced in Subsection~\ref{subsec:4_1}.
Let ${\mathcal T}$ be the free abelian group generated by $T_1,\ldots,T_N$ which
acts on ${\mathcal P}(\m)$.
% in the same way as the group ${\mathcal T}_k$ % in (\ref{eq:apr6_1})
% does for each $k$.
We def\/ine its actions on the set $\Jb (\m)$.
Given any $h = \sum_{k=1}^N n_k T_k \in {\mathcal T}$
% $\eta_j(h, J) = ({J'}_i)_{i \in \Z} \in \J (m,p)$ with
% ${J'}_i = J_{i} + \sum_{k=1}^N n_k \min (j,k)$.
and $\boJ = (J^{(j)}_i)_{i \in \Z, 1 \leq j \leq N} \in \Jb (\m)$
we set
\begin{displaymath}
h \cdot \boJ = \left( J_{i}^{(j)} + \sum_{k=1}^N n_k \min (j,k) \right)_{i \in \Z, 1 \leq j \leq N}.
\end{displaymath}

The direct scattering transform $\Phi$ ((3.15) in~\cite{KTT}) is given as follows:
\begin{equation}\label{eq:pj}
\begin{split}
\Phi: \quad &\Pth(\m) \longrightarrow \;\Z \times \Pth_+(\m)\;\;
\longrightarrow \;\;\quad \Jb(\m) \;\;\;\;\longrightarrow
\;\; {{\mathcal A} \backslash \Jb (\m)}\\
&\;\;\;p \;\quad \longmapsto \;\;\;\;(d, p_+) \;\;\quad\longmapsto
\quad \iota(\boJ)+d \;\;\longmapsto \;\;[\iota(\boJ)+d]_{\mathcal A}
\end{split}
\end{equation}
Here the second map is given by $\phi$ together with the map $\iota$ {(\ref{eq:iota})}, where
for any $d \in \Z$ and $\boJ = ((J^{(j)}_i)_{1 \le i \le m_{j}})_{1 \leq j \leq N} \in \Rig(\m)$ we write
{$\iota (\boJ) + d = (J^{(j)}_i + d)_{i \in \Z, 1 \leq j \leq N} \in \Jb(\m)$}.
{As in Subsection~\ref{subsec:4_2},}
suppose there are two possible ways to write $p$ as  $p =(d T_1) \cdot p_+ = (d' T_1) \cdot p'_+$.
{Then one generally has $\iota(\boJ')+d' \ne \iota(\boJ)+d$ in $\Jb(\m)$.}
However it can be shown that the relation $[\iota(\boJ')+d']_{\mathcal A} = [\iota(\boJ)+d]_{\mathcal A}$
{holds}, permitting the map $\Phi$ to be well-def\/ined.
\begin{example}\label{ex:oct16_2}
Let $\boJ$ and $\boJ'$ be the ones in Example \ref{ex:oct16_1}.
Then we have
$(\sigma_1 + \sigma_2) \cdot (\iota (\boJ) + d) = \iota (\boJ') + d'$ with $d=0$, $d'=6$.
Hence for the $p$ and $\m$ in that example,
{$\Phi (p) = [\iota (\boJ) + d]_{\mathcal A} = [\iota (\boJ') + d']_{\mathcal A}$}.
\end{example}

It is easy to see that the actions of ${\mathcal A}$ and ${\mathcal T}$ on $\Jb(\m)$ is commutative.
Hence by Lemma \ref{lem:jan29_1} there is a natural action of ${\mathcal T}$ on ${\mathcal A} \backslash \Jb (\m)$.

The main result of \cite{KTT} is as follows.
\begin{proposition}[\protect{\cite[Theorem 3.12]{KTT}}]\label{prop:ktt}
The map $\Phi$ is a bijection
between ${\mathcal P}(\m)$ and ${\mathcal A} \backslash \Jb (\m)$
that is commutative with the action of ${\mathcal T}$.
\end{proposition}

In particular one has $|{\mathcal P}(\m)| = |{\mathcal A} \backslash \Jb (\m)|$.
This quantity is written as $\Omega (\m)$ in (\ref{eq:oct10_1}) where the $N \times N$ matrix
$F$ is set to be the $F^{(\boa)}$ in (\ref{eq:may15_1}) with (\ref{eq:may15_2}), (\ref{eq:may26_1}) and
$\boa = (1,\ldots,1)$.
The level set itself is written as $\J (\m) {:= {\mathcal A} \backslash \Jb (\m)}$ in (\ref{eq:jv}) where the
$\gamma \times \gamma$ matrix $A$ is def\/ined by $A = (A_{j\alpha,k\beta})_{1 \leq j,k \leq N, 1 \leq \alpha \leq m_j, 1 \leq \beta \leq m_k}$
with
\begin{math}
A_{j\alpha,k\beta} = \delta_{j,k} \delta_{\alpha,\beta} (p_j + m_j) + 2 \min (j,k) - \delta_{j,k}.
\end{math}

%%%%%%%%%%%%%%%%%%%%%%%%%%%%%%%%%%%%%%%%%%%%%%%%%%%%%%%%%%%%%%%%%%%%%%%%%%%%%%%%%
\subsection{Construction of the invariant tori}\label{subsec:4_4}
Let $\alpha$ be a common divisor of $m$ and $p$.
As a subset of $\Jb (m,p)$ %in (\ref{eq:jcomp})
we def\/ine
\begin{displaymath}
{\Jb}^{(\alpha)} (m,p) = \{
J \in {\Jb} (m/\alpha,p/\alpha) \mid
J \notin {\Jb} (m/\alpha',p/\alpha') \, \mbox{for every common divisor} \, \alpha' > \alpha \}.
\end{displaymath}
Choose a sequence $\boa = (\alpha_j)_{1 \leq j \leq N}$ with
each $\alpha_j$ being a common divisor of $m_j$ and $p_j$.
We def\/ine
$\Jb^{(\boa)} (\m) = \prod_{j=1}^N {\Jb}^{(\alpha_j)} (m_j,p_j)$.
This $\boa$ represents the symmetry of the system that depends on the initial conditions.
As a subset of $\Jb (\m)$ it is invariant under the actions of ${\mathcal A}$ and ${\mathcal T}$.
{Let
% $\J^{(\boa)} (\m) = {\mathcal A} \backslash \Jb^{(\boa)} (\m)$ and
${\mathcal P}^{(\boa)}(\m) = \Phi^{-1}(\Jb^{(\boa)} (\m))$.}
As a corollary of Proposition~\ref{prop:ktt} we have the following.
\begin{corollary}\label{coro:ktt}
The map $\Phi$ is a bijection
between ${\mathcal P}^{(\boa)}(\m)$ and ${\mathcal A} \backslash
\Jb^{(\boa)} (\m)$
that is commutative with the action of ${\mathcal T}$.
\end{corollary}

Let $X^{(\boa)} = X_1 \times X_2^{(\boa)}$  be the set introduced in (\ref{eq:apr22_2}).
In what follows
we identify the group~$H_2$ in Subsection \ref{subsec:3_1} with the group ${\mathcal A}$ by $s_i = \sigma_i$,
and set $p_j$ and $B_{j,k}$ as in (\ref{eq:may15_2}) and (\ref{eq:may26_1}).

\begin{lemma}\label{lem:may14_1}
There is a bijection between $\Jb^{(\boa)} (\m)$ and $X^{(\boa)}$ that is commutative with the action of ${\mathcal A}$.
\end{lemma}

\begin{proof}
For each {$\boJ = \big(J^{(j)}_i\big)_{i \in \Z, 1 \leq j \leq N} \in \Jb^{(\boa)} (\m)$}
let
\begin{displaymath}
\bom = \big(J^{(j)}_1\big)_{1 \leq j \leq N}, \qquad \bol = \big(J^{(j)}_i-J^{(j)}_1\big)_{i \in \Z, 1 \leq j \leq N}.
\end{displaymath}
Then the map $\Jb^{(\boa)} (\m) \ni \boJ \mapsto (\bom, \bol) \in X^{(\boa)}$ yields the desired bijection.
\end{proof}

By this lemma we can identify the set $\tilde{X} = X_1 \times X_2$ in Subsection \ref{subsec:3_1}
with the set
$\Jb (\m)$.

\begin{example}\label{ex:oct16_3}
Let $\boJ$, $\boJ'$, $d$, $d'$ be the ones in Examples \ref{ex:oct16_1} and \ref{ex:oct16_2}.
Then
$\iota (\boJ) + d$ and $\iota (\boJ') + d'$ are elements of $\Jb^{(\boa)} (\m)$ with
$\m$ in those examples, and $\boa = (\alpha_1, \alpha_2, \alpha_3) = (3,1,1)$.
The above map sends $\iota (\boJ) + d$ to the $(\bom, \bol) \in X^{(\boa)}$ in Example \ref{ex:oct15_4}.
Write the image of $\iota (\boJ') + d'$ as $(\bom', \bol') \in X^{(\boa)}$.
Then $\bom' = (8,7,6) \in X_1 = \Z^3$, and $\bol'$ is the one labeled by $(a,b,c) = (4,8,5)$.
By using the formula in Example \ref{ex:oct15_3} one can show that $(s_1 + s_2) \cdot (\bom, \bol) = (\bom', \bol')$.
Compared with Example \ref{ex:oct16_2} this shows the commutativity of the action of ${\mathcal A}$ with
the bijection in Lemma~\ref{lem:may14_1}.
\end{example}

By Lemmas \ref{lem:may14_1} and \ref{lem:apr8_1} (to appear in the appendix)
there is a natural action of ${\mathcal T}$ on $X^{(\boa)}$ that is commutative with the action of ${\mathcal A}$.
Also one has a bijection between {${\mathcal A} \backslash \Jb^{(\boa)} (\m)$} and
${\mathcal A} \backslash X^{(\boa)}$ that is commutative with the action of ${\mathcal T}$.
Now by Corollary \ref{coro:ktt} we have the following.
\begin{proposition}\label{prop:may14_2}
There is a bijection between ${\mathcal P}^{(\boa)}(\m)$ and ${\mathcal A} \backslash X^{(\boa)}$
that is commutative with the action of ${\mathcal T}$.
\end{proposition}
{We denote by $\Psi$ the map representing this bijection.
Then for any $p \in {\mathcal P}^{(\boa)}(\m)$ there exists some $x \in X^{(\boa)}$ such that
$\Psi (p) =  [x]_{\mathcal A}$.
Now as a corollary of Proposition \ref{prop:may14_2} we have the following.}
\begin{proposition}\label{prop:apr22_3}
Given $p \in {\mathcal P}^{(\boa)}(\m)$ and $x \in X^{(\boa)}$ as above,
there is a bijection between ${\mathcal T} \cdot p (\subset {\mathcal P}^{(\boa)}(\m))$
and ${\mathcal T} \cdot  [x]_{\mathcal A} (\subset {\mathcal A} \backslash X^{(\boa)})$
that is commutative with the action of ${\mathcal T}$.
\end{proposition}

The action of the abelian group ${\mathcal T}$ on the set $X^{(\boa)} = X_1 \times X_2^{(\boa)}$
keeps $X_2^{(\boa)}$ untouched.
Thus this action is essentially def\/ined on the set $X_1 = \Z^N$.
By Lemma \ref{lem:apr8_1} it is uniquely determined from the action of ${\mathcal T}$ on the set ${\mathcal P}(\m)$
in Subsection~\ref{subsec:4_3}.
An explicit description of this action is as follows.
For each $h = \sum_{k=1}^N n_k T_k \in {\mathcal T}$ and $\bom = (\omega^{(j)})_{1 \leq j \leq N} \in X_1$
we set $h \cdot \bom = \bom + \sum_{k=1}^N n_k \boldsymbol{h}_k$ where
$\boldsymbol{h}_k = (\min(j,k))_{1 \leq j \leq N} \in X_1$.

\begin{lemma}\label{lem:oct18_2}
The action of the group ${\mathcal T}$ on the set $X_1 = \Z^N$ is transitive.
\end{lemma}
\begin{proof}
Let $\boldsymbol{\omega}$, $\boldsymbol{\omega'}$ be any two elements of $\Z^N$,
thought of as column vectors.
We def\/ine the $N \times N$ matrix $M = (\boldsymbol{h}_1,\ldots,\boldsymbol{h}_N)$.
It is easy to see that $\det M = 1$.
Hence by $\boldsymbol{n} =M^{-1} (\boldsymbol{\omega'} - \boldsymbol{\omega})$
we obtain a column vector $\boldsymbol{n} = {}^t (n_1,\ldots,n_N)$ with integer entries.
Then we have $\boldsymbol{\omega'} =(n_1 T_1 + n_2 T_2 + \cdots + n_N T_N) \cdot \boldsymbol{\omega}$.
\end{proof}
\begin{remark}\label{rem:may8_1}
For the generalized case in Remark \ref{rem:mar12_1}
we def\/ine ${\mathcal T}$ to be the free abelian group
generated by $T_1, T_{j_1+1},\ldots, T_{j_{s-1}+1}$.
Then one can show that the action of ${\mathcal T}$ on the set $X_1=\Z^s$ is transitive
in the same way as above.
\end{remark}

By Lemma \ref{lem:oct18_2} one can adopt ${\mathcal T}$ as the abelian group $H_1$ in Subsection~\ref{subsec:3_1}.
Given any $p \in {\mathcal P}^{(\boa)}(\m)$, a path of soliton content $\m$ and with the symmetry $\boa$,
one can obtain the matrix~$F^{(\boa)}$ by (\ref{eq:may15_1}),
(\ref{eq:may15_2}) and (\ref{eq:may26_1}).
By Propositions \ref{prop:apr22_1} and \ref{prop:apr22_3} we f\/inally obtain our second main result.
\begin{theorem}\label{th:main}
There is a bijection between ${\mathcal T} \cdot p$ and $F^{(\boa)} \Z^N \backslash \Z^N$
that is commutative with the action of ${\mathcal T}$.
\end{theorem}

As we have shown at the end of Subsection~\ref{subsec:3_3} the $F^{(\boa)} \Z^N \backslash \Z^N$ is an $N$-dimensional discrete torus.
By the considerations given just above Lemma \ref{lem:oct18_2} one also knows that
each time evolution~$T_k$ yields a straight motion on the torus with the ``velocity'' vector
$\boldsymbol{h}_k = (\min(j,k))_{1 \leq j \leq N}$.

\begin{example}\label{ex:oct16_4}
For $p=121122111212211222121111$ and
\begin{equation*}
F^{(\boa)} =
\begin{pmatrix}
6 & 4 & 2 \\
2 & 14 & 4 \\
2 & 8 & 10
\end{pmatrix}
,
\end{equation*}
there is a bijection between ${\mathcal T} \cdot p$ and $F^{(\boa)} \Z^3 \backslash \Z^3$.
This is a consequence of Examples \ref{ex:oct15_4}, \ref{ex:oct16_1} and~\ref{ex:oct16_3}.
\end{example}

\section{Periods and the level set structure}\label{sec:5}
\subsection{Dynamical periods}\label{subsec:5_1}

Given $p \in {\mathcal P}^{(\boa)}(\m)$, a path of soliton content $\m$ and with the symmetry $\boa$,
there is a smallest positive integer $n_k$ such that $(n_k T_k) \cdot p = p$ for each $k$.
It is called the \textit{dynamical period} of $p$ associated with the time evolution $T_k$.
Since the time evolutions are mapped to straight motions on the torus $F^{(\boa)} \Z^N \backslash \Z^N$,
it is easy to obtain an explicit formula for $n_k$.

We f\/ix a bijection $\widehat{\Psi}$ from ${\mathcal T} \cdot p$ to $F^{(\boa)} \Z^N \backslash \Z^N$
whose existence was ensured by Theorem \ref{th:main}.
Then since $p$ lies in ${\mathcal T} \cdot p$ there exists $\bom \in \Z^N$ such that $\widehat{\Psi} (p) =  [\bom]_{F^{(\boa)} \Z^N}$.
By the def\/inition of the action of ${\mathcal T}$ on $F^{(\boa)} \Z^N \backslash \Z^N$ we have
\begin{displaymath}
(n_k T_k) \cdot \widehat{\Psi} (p) = [(n_k T_k) \cdot \bom ]_{F^{(\boa)} \Z^N}
= [\bom + n_k \boldsymbol{h}_k ]_{F^{(\boa)} \Z^N}.
\end{displaymath}
On the other hand since $\widehat{\Psi}$ is commutative with the action of ${\mathcal T}$ we have
\begin{displaymath}
(n_k T_k) \cdot \widehat{\Psi} (p) = \widehat{\Psi} ((n_k T_k) \cdot p) =  \widehat{\Psi} (p)= [\bom]_{F^{(\boa)} \Z^N}.
\end{displaymath}
Thus $n_k \boldsymbol{h}_k$ must lie in the $F^{(\boa)} \Z^N$-orbit of $0$, implying that
$n_k$ is def\/ined as the smallest positive integer satisfying $n_k \boldsymbol{h}_k \in F^{(\boa)} \Z^N$.
This condition is essentially the same one given in~(4.32) of~\cite{KTT} which yields an explicit formula for this quantity ((4.26) of~\cite{KTT}).
Note that our formulation uses no Bethe ansatz considerations to obtain this result.
\begin{example}\label{ex:oct16_5}
For $p=121122111212211222121111$ one can observe
the dynamical periods $n_1 = 24$, $n_2 = 48$ and $n_3=72$ by directly applying the time evolutions
in Subsection~\ref{subsec:4_1}.
By using the above considerations they are also calculated as
\begin{equation*}
n_k = {\rm L.C.M.} \left( \frac{\det  F^{(\boa)}}{\det  F^{(\boa)}[1]},
\frac{\det  F^{(\boa)}}{\det  F^{(\boa)}[2]}, \frac{\det  F^{(\boa)}}{\det  F^{(\boa)}[3]} \right),
\end{equation*}
where $F^{(\boa)}$ is the one in Example \ref{ex:oct16_4} and
$F^{(\boa)}[i]$ is the matrix obtained from $F^{(\boa)}$ by replacing its $i$-th column by $\boldsymbol{h}_k$.
Here L.C.M.~stands for the least common multiple.
For instance, we have
\begin{equation*}
n_3 = {\rm L.C.M.} \left( 24,
\frac{72}{5}, \frac{72}{17} \right) = 72.
\end{equation*}
\end{example}

\subsection{Decomposition of the level set}\label{subsec:5_2}
Each element $\lambda = (\lambda_i)_{i \in \Z}$ of the set $\Lambda (m,p)$
is specif\/ied by
a non-decreasing sequence of integers $(0 \leq) \lambda_2 \leq \lambda_3 \leq \dots \leq \lambda_m (\leq p)$,
or a partition $(\lambda_m, \ldots, \lambda_2)$.
Hence the cardinality of $\Lambda (m,p)$ is given by the binomial coef\/f\/icient
\begin{math}
|\Lambda (m,p)| = { p+m-1 \choose m-1 }.
\end{math}
Recall the decomposition of the set $\Lambda (m,p) = \bigsqcup_{\alpha} \Lambda^{(\alpha)} (m,p)$,
where $\alpha$ runs over every common divisor of $m$ and $p$.
Thus we have the relation
\begin{equation*}
{ p+m-1 \choose m-1 } = \sum_\alpha |\Lambda^{(\alpha)} (m,p)|.
\end{equation*}
The formula (\ref{eq:oct10_1}) for the cardinality of
the whole level set with soliton content $\m$ is now written as
\begin{gather}
\Omega(\m) =
(\det F)\prod_{1 \leq j \leq N} \frac{1}{m_j}
\binom{p_j + m_j - 1}{m_j - 1} \nonumber\\
\phantom{\Omega(\m)}{}
=
\sum_{\alpha_1} \cdots \sum_{\alpha_N} \big(\det F^{(\boa)}\big)
\prod_{1 \leq j \leq N} \frac{|\Lambda^{(\alpha_j)} (m_j,p_j)|}{m_j/\alpha_j}.
\label{eq:oct24_1}
\end{gather}
This expression for the volume formula shows the decomposition of the level set (\ref{eq:jv}) into the invariant tori:
For each $\boa = (\alpha_j)_{1 \leq j \leq N}$
we have the tori $F^{(\boa)} \Z^N \backslash \Z^N$ with the multiplicity
$\prod_{1 \leq j \leq N} \frac{|\Lambda^{(\alpha_j)} (m_j,p_j)|}{m_j/\alpha_j}$.
See \cite[Lemmas 3.1 and 3.2]{KT} for a proof of this statement.
While  the smallest torus is given by $\boa = (\alpha_j)_{1 \leq j \leq N}$ with every $\alpha_j$ being the greatest common divisor
of~$m_j$ and~$p_j$,
the largest torus is given by  $\boa = (1)_{1 \leq j \leq N}$ when $p_N > 0$, or $\boa = (1,\ldots,1,m_N)$ when $p_N =0$.

One can check that this expression for the
multiplicity is indeed an integer.
Let $s$ be the generator of the cyclic group ${\mathcal C}_m$ that acts on $\Lambda^{(\alpha)} (m,p)$.
For each $\lambda \in \Lambda^{(\alpha)} (m,p)$ the $m/\alpha$ elements $(k s) \cdot \lambda \, (0 \leq k \leq m/\alpha-1)$ are all distinct
and hence $|\Lambda^{(\alpha)} (m,p)|$ is divisible by $m/\alpha$.

An expression for the quantity $|\Lambda^{(\alpha)} (m,p)|$ is given as follows.
Let $\mu (n)$ be the M\"obius function of number theory~\cite{S}; that is,
$\mu (1)=1, \mu (n)=0$ if $n$ is divisible by the square of an integer greater than one,
and $\mu (n) = (-1)^r$ if $n$ is the product of $r$ distinct primes.
Then we have
\begin{equation}\label{eq:oct16_7}
|\Lambda^{(\alpha)} (m,p)| = \sum_\beta \mu(\beta/\alpha) {\frac{p+m}{\beta} -1 \choose \frac{m}{\beta} -1},
\end{equation}
where $\beta$ runs over every common divisor of $m$ and $p$ that is a multiple of $\alpha$.

\begin{example}\label{ex:oct16_6}
For $L=24$ and $\m = (m_1, m_2, m_3) = (3,2,1)$ we have $(p_1, p_2, p_3) = (12,6,4)$.
The former expression of (\ref{eq:oct24_1}) gives
\begin{gather*}
\Omega(\m) = (\det F)
\frac{1}{m_1} \binom{p_1 + m_1 - 1}{m_1 - 1}
\frac{1}{m_2} \binom{p_2 + m_2 - 1}{m_2 - 1}
\frac{1}{m_3} \binom{p_3 + m_3 - 1}{m_3 - 1}\\
\phantom{\Omega(\m)}{}
= 1728 \cdot \frac13 \binom{14}{2} \cdot \frac12 \binom{7}{1} =183456,
\end{gather*}
where $F = F^{(\boa_1)}$ is given below.
Let $\boa_1 = (1,1,1)$, $\boa_2 = (1,2,1)$, $\boa_3 = (3,1,1)$, $\boa_4 = (3,2,1)$.
Then
\begin{gather*}
F^{(\boa_1)} =
\begin{pmatrix}
18 & 4 & 2 \\
6 & 14 & 4 \\
6 & 8 & 10
\end{pmatrix}
,\qquad
F^{(\boa_2)} =
\begin{pmatrix}
18 & 2 & 2 \\
6 & 7 & 4 \\
6 & 4 & 10
\end{pmatrix}
,\\
F^{(\boa_3)} =
\begin{pmatrix}
6 & 4 & 2 \\
2 & 14 & 4 \\
2 & 8 & 10
\end{pmatrix}
,\qquad
F^{(\boa_4)} =
\begin{pmatrix}
6 & 2 & 2 \\
2 & 7 & 4 \\
2 & 4 & 10
\end{pmatrix}.
\end{gather*}
By using (\ref{eq:oct16_7}) one has
$|\Lambda^{(1)}(3,12)| = 90$, $|\Lambda^{(3)}(3,12)| = 1$, $|\Lambda^{(1)}(2,6)| = 6$ and $|\Lambda^{(2)}(2,6)| = 1$.
See Example \ref{ex:oct15_2} for the decomposition of the set $\Lambda (m,p)$ in the present example.
Now the latter expression of (\ref{eq:oct24_1}) gives
\begin{equation*}
\Omega(\m) = 90 \det F^{(\boa_1)} + 30 \det F^{(\boa_2)} + 3 \det F^{(\boa_3)} + \det F^{(\boa_4)}=183456.
\end{equation*}
This enumeration ref\/lects the following decomposition of the level set into invariant tori:
\begin{equation*}
{\mathcal J} (\m) = 90 (F^{(\boa_1)} \Z^3 \backslash \Z^3) \sqcup
30 (F^{(\boa_2)} \Z^3 \backslash \Z^3) \sqcup
3 (F^{(\boa_3)} \Z^3 \backslash \Z^3) \sqcup (F^{(\boa_4)} \Z^3 \backslash \Z^3).
\end{equation*}
The ${\mathcal T} \cdot p \cong F^{(\boa)} \Z^3 \backslash \Z^3$ in Example \ref{ex:oct16_4}
is one of the three $F^{(\boa_3)} \Z^3 \backslash \Z^3$s in this ${\mathcal J} (\m)$.
\end{example}

\appendix
\section{Two elementary lemmas}\label{app:A}
The following lemma is used in Subsection \ref{subsec:3_3}
\begin{lemma}\label{lem:apr8_2}
Let $G, H$ be groups and $X$ be a set.
Suppose there are commutative actions of $G$ and $H$ on $X$.
Denote by $\varGamma$ $($resp.~$\varGamma')$ the action of $G$ $($resp.~$H)$ on $X$, and by $\rho_{\varGamma}$ $($resp.~$\rho_{\varGamma'})$
its permutation representation.
Then:
\begin{enumerate}\itemsep=0pt
\item The actions of $\rho_{\varGamma} (G)$ and $H$ on $X$ are commutative.
\item There is a bijection between $G \backslash X$ and $\rho_{\varGamma} (G) \backslash X$ that is commutative with the
action of~$H$.
\item For all $x \in X$ there is a bijection between $H \cdot [x]_G$ and $\rho_{\varGamma'}(H) \cdot  [x]_{\rho_{\varGamma}(G)}$
that is commutative with the action of $H$.
\end{enumerate}
\end{lemma}
\begin{proof}
$(i)$ For all $g \in G, h \in H$ and $x \in X$ we have $h \cdot (\rho_{\varGamma} (g) \cdot x) = h \cdot \varGamma (g, x)
= \varGamma (g, h \cdot x) = \rho_{\varGamma} (g) \cdot (h \cdot x)$.
$(ii)$ Given any $p \in G \backslash X$ it can be written as $p = [x]_G$ with some $x \in X$.
Def\/ine $\psi: G \backslash X \rightarrow \rho_{\varGamma} (G) \backslash X$ by the relation $\psi (p) = [x]_{\rho_{\varGamma} (G)}$.
It is easy to see that this map is well-def\/ined and yields the desired bijection.
$(iii)$ The bijection is given from that in $(ii)$ by restricting its domain to $H \cdot [x]_G$.
\end{proof}

The following lemma is used in Subsection~\ref{subsec:4_4}
\begin{lemma}\label{lem:apr8_1}
Let $G$ be a group and $X_1$, $X_2$ be sets.
Suppose there are actions of $G$ on~$X_1$ and~$X_2$, and there is a bijection $\phi : X_1 \rightarrow X_2$ that is
commutative with the actions of~$G$.
Let~$H$ be another group, and suppose there is an action of $H$ on $X_1$ that is commutative with the action of $G$.
Then:
\begin{enumerate}\itemsep=0pt
\item There is a unique action of $H$ on $X_2$ that is commutative with $\phi$, and commutative with
the action of $G$ on $X_2$.
\item There is a bijection between $G \backslash X_1$ and  $G \backslash X_2$ that is commutative with the action of $H$.
\end{enumerate}
\end{lemma}
\begin{proof}
$(i)$ For all $h \in H$ and $x \in X_2$
the desired action is uniquely determined by $h \cdot x = \phi (h \cdot \phi^{-1} (x))$.
$(ii)$
Given any $p \in G \backslash X_1$ it can be written as $p = [x]_G$ with some $x \in X_1$.
Def\/ine $\bar{\phi}: G \backslash X_1 \rightarrow G \backslash X_2$ by the relation $\bar{\phi} (p) = [\phi (x)]_G$.
It is easy to see that this map is well-def\/ined and yields the desired bijection.
\end{proof}

\section[An algorithm for Kerov-Kirillov-Reshetikhin map]{An algorithm for Kerov--Kirillov--Reshetikhin map}\label{app:B}

We present an algorithm for the map of Kerov--Kirillov--Reshetikhin
based on~\cite{Takagi}.

Let $p = b_1 \ldots b_L$ be a ballot sequence of letters $1$ and $2$.
We def\/ine $M_1$, $M_2$ to be a pair of subsets of $\{ 1, \ldots, L \}$ that is specif\/ied by
the following condition:
The integer $j$ lies in $M_1$ (resp.~$M_2$) if and only if $b_j = 1$ and $b_{j+1} = 2$
(resp.~$b_j = 2$ and $b_{j+1} = 1$) .
Here we interpret $b_{L+1}=1$.

We def\/ine ${\rm Op}_1$, ${\rm Op}_2$ to be a pair of operators acting
on f\/inite sets of distinct integers:
Given any such set $M$, the ${\rm Op}_1$ (resp.~${\rm Op}_2$) replaces its $i$-th smallest element, say~$x$, by
$x - (2i-1)$ (resp.~$x - 2i$) from $i=1$ to $i=|M|$.

Let $\overline{\varphi} = \{ \}$ be an empty data set.
We repeat the following procedure until we have $M_1 = M_2 = \varnothing$.
\begin{enumerate}\itemsep=0pt
\item Let $i=0$.
\item While $M_1 \cap M_2 = \varnothing$, apply ${\rm Op}_1$ to $M_1$ and ${\rm Op}_2$ to $M_2$, and
replace $i$ by $i+1$.
\item When $M_1 \cap M_2 \ne \varnothing$ is attained, append $\{ M_1 \cap M_2 , i\}$ to $\overline{\varphi}$,
and replace $M_1$ by $M_1 \setminus (M_1 \cap M_2)$ and $M_2$ by $M_2 \setminus (M_1 \cap M_2)$.
\end{enumerate}
In the above procedure the multiplicity of the
elements should be respected.
For instance $\{1,2,2,3\} \cap \{2,2,4,5\}$ is
equal to $\{2,2\}$, not to $\{2\}$.
And $\{1,2,2,3\} \setminus \{2,3\}$ is
equal to $\{1,2\}$, not to $\{1\}$.
At the end we obtain such type of data
$\overline{\varphi}=\{ \{S_1,i_1\}, \{S_2,i_2\},\ldots,\{S_s,i_s\} \}$
for some~$s$, where $S_a$ are sets of integers and~$i_a$ are positive integers.

Now we have ${\mathcal H} = \{ j_1, \ldots, j_s \}$
with $j_1=i_1, j_2=i_1+i_2, \ldots, j_s = i_1 + \cdots + i_s$,
and $\m = (m_{j_1}, \ldots, m_{j_s}) = (|S_1|, \ldots, |S_s|)$.
For the riggings $\boJ$, each $(J^{(j_k)}_i)_{1 \le i \le m_{j_k}}$ is obtained from $S_k$ by
ordering its elements in increasing order.

\begin{example}
Consider the path $p=11112222111112221112111222111222212222$.
Then $M_1 = \{ 4,13,19,23,29, 34 \}$ and $M_2 = \{ 8, 16, 20, 26, 33, 38 \}$.
By applying procedure $(ii)$ once,
we obtain
$M_1 = \{ 3,10,14,16,20, 23 \}$ and $M_2 = \{ 6, 12, 14, 18, 23, 26 \}$.
Then by procedure $(iii)$ one has $\overline{\varphi}=\{ \{ \{ 14, 23 \},1\} \}$,
$M_1 = \{ 3,10,16,20 \}$, and $M_2 = \{ 6, 12, 18, 26 \}$.

In the second turn of the algorithm,
we obtain
$M_1 = \{ 1,4,6,6 \}$ and $M_2 = \{ 2,4,6,10 \}$ after applying procedure $(ii)$ twice.
Thus one has $\overline{\varphi}=\{ \{ \{ 14, 23 \},1\}, \{ \{ 4, 6 \},2\} \}$,
$M_1 = \{ 1,6 \}$, and $M_2 = \{ 2, 10 \}$.

After two more turns,
one has $\overline{\varphi}=\{ \{ \{ 14, 23 \},1\}, \{ \{ 4, 6 \},2\}, \{ \{ 0 \},1\}, \{ \{ 0 \},3\} \}$
and $M_1 =  M_2 = \varnothing$.
Hence we have ${\mathcal H} = \{ 1,3,4,7 \}$, $\m = (m_1,m_3,m_4,m_7)= (2,2,1,1)$ and
\begin{gather*}
\boJ  = \big(\big(J^{(1)}_1,J^{(1)}_2\big),\big(J^{(3)}_1,J^{(3)}_2\big),\big(J^{(4)}_1\big),\big(J^{(7)}_1\big)\big) = ((14,23),(4,6),(0),(0)).
\end{gather*}

\end{example}

\subsection*{Acknowledgements}
The author thanks Atsuo Kuniba for
valuable discussions.

\pdfbookmark[1]{References}{ref}
\LastPageEnding

\end{document}